\documentclass[aps,prd,reprint,preprintnumbers,superscriptaddress,amsmath,amssymb,floatfix]{revtex4-2}

\usepackage{graphicx}% Include figure files
\usepackage{epsfig}
\usepackage{dcolumn}% Align table columns on decimal point
\usepackage{bm}% bold math
\usepackage{multirow}
\usepackage[colorlinks,linkcolor=red,anchorcolor=green,citecolor=blue,CJKbookmarks=True]{hyperref}

\begin{document}
\title{$\eta$-glueball mixing from $N_f=2$ lattice QCD}

\author{Xiangyu Jiang}
\email{jiangxiangyu@ihep.ac.cn}
\affiliation{Institute of High Energy Physics, Chinese Academy of Sciences, Beijing 100049, People's Republic of China}
\affiliation{School of Physical Sciences, University of Chinese Academy of Sciences, Beijing 100049, People's Republic of China}

\author{Wei Sun}
\email{sunwei@ihep.ac.cn}
\affiliation{Institute of High Energy Physics, Chinese Academy of Sciences, Beijing 100049, People's Republic of China}

\author{Feiyu Chen}
\affiliation{Institute of High Energy Physics, Chinese Academy of Sciences, Beijing 100049, People's Republic of China}
\affiliation{School of Physical Sciences, University of Chinese Academy of Sciences, Beijing 100049, People's Republic of China}

\author{Ying Chen}
\email{cheny@ihep.ac.cn}
\affiliation{Institute of High Energy Physics, Chinese Academy of Sciences, Beijing 100049, People's Republic of China}
\affiliation{School of Physical Sciences, University of Chinese Academy of Sciences, Beijing 100049, People's Republic of China}

\author{Ming Gong}
\affiliation{Institute of High Energy Physics, Chinese Academy of Sciences, Beijing 100049, People's Republic of China}
\affiliation{School of Physical Sciences, University of Chinese Academy of Sciences, Beijing 100049, People's Republic of China}

\author{Zhaofeng Liu}
\affiliation{Institute of High Energy Physics, Chinese Academy of Sciences, Beijing 100049, People's Republic of China}
\affiliation{School of Physical Sciences, University of Chinese Academy of Sciences, Beijing 100049, People's Republic of China}
\affiliation{Center for High Energy Physics, Peking University, Beijing 100871, People's Republic of China}

\author{Renqiang Zhang}
\affiliation{Institute of High Energy Physics, Chinese Academy of Sciences, Beijing 100049, People's Republic of China}
\affiliation{School of Physical Sciences, University of Chinese Academy of Sciences, Beijing 100049, People's Republic of China}

\begin{abstract}
    We perform the first lattice study on the mixing of the isoscalar pseudoscalar meson $\eta$ and the pseudoscalar glueball $G$ in the $N_f=2$ QCD at the pion mass $m_\pi\approx 350$ MeV. The $\eta$ mass is determined to be $m_\eta=714(6)(16)$ MeV. Through the Witten-Veneziano relation, this value can be matched to a mass value of $\sim 981$ MeV for the $\mathrm{SU(3)}$ counterpart of $\eta$. Based on a large gauge ensemble, the $\eta-G$ mixing energy and the mixing angle are determined to be $|x|=107(15)(2)$ MeV and $|\theta|=3.46(46)^\circ$ from the $\eta-G$ correlators that are calculated using the distillation method. We conclude that the $\eta-G$ mixing is tiny and the topology induced interaction contributes most of $\eta$ mass owing to the QCD $\mathrm{U_A(1)}$ anomaly.
\end{abstract}
%\pacs{12.38.Aw, 12.40.Yx}
\maketitle
\section{Introduction}
Chiral symmetry is an intrinsic symmetry of quantum chromodynamics (QCD) in the massless limit of quarks and is spontaneously broken due to the nonzero quark condensate. The spontaneous symmetry breaking (SSB) of the $\mathrm{SU_L(3)}\times \mathrm{SU_R(3)}$ into the flavor $\mathrm{SU(3)}$ generates eight Goldstone particles sorted into the pseudoscalar octet composed of $\{\pi, K, \eta_8\}$ which are slightly massive due to the small $u,d,s$ quark masses. If SSB also applies to the $\mathrm{U_L(1)}\times \mathrm{U_R(1)}$ sector of the chiral symmetry, it breaks into $\mathrm{U_V(1)}$ that corresponds to the baryon number conservation expects the existence of an additional Goldstone particle, namely, a light flavor singlet pseudoscalar meson. The $\eta'$ meson is predominantly a flavor singlet but is too massive to be taken as a candidate for this Goldstone particle. The $\eta'$ mass puzzle has a direct connection with the QCD $\mathrm{U_A(1)}$ anomaly that the anomalous gluonic term, the topological charge density, breaks the conservation of the flavor singlet axial current even in the chiral limit. Even though the anomalous axial vector relation can be written as the divergence of a gauge variant axial vector, which is zero and implies a $\mathrm{U_A(1)}$ symmetry, Kogut and Susskind~\cite{Kogut:1974kt} pointed out that its spontaneous breaking generates a massless mode that violates the Gell-Mann-Okubo relation and then renders $\eta'$ more massive. With respect to the nontrivial topology of QCD vacuum, Witten~\cite{Witten:1978bc} and Veneziano~\cite{Veneziano:1979ec} proposed a mechanism for the original of $\eta'$ mass that the nonperturbative coupling of the topological charge density and the flavor singlet pseudoscalar induces a  self energy correction $m_0^2$, which is proportional to the topological susceptibility $\chi$ of gauge fields. Using the physical mass of $\eta'$, the value of $\chi$ is estimated to be around $(180~\mathrm{MeV})^4$, which is supported by lattice QCD calculations~\cite{Cichy:2015jra}.  Another interesting property of $\eta'$ is its large production rate in the $J/\psi$ radiative decays~\cite{Zyla:2020zbs}, which are abundant in gluons and are expected to favor the production of glueballs. This observation along with the original mechanism of $\eta'$ mass manifests the strong coupling of $\eta'$ to gluons and thereby prompts the conjecture that $\eta'$ may mix substantially with pseudoscalar glueballs since they have the same quantum number. The KLOE Collaboration analyzed the processes $\phi\to \gamma \eta$ and $\phi\to \gamma\eta'$ and found that the $\eta'$-glueball mixing might be required and the mixing angle can be as large as $(22\pm 3)^\circ$~\cite{KLOE:2006guu}. In contrast, another phenomenological analysis of the KLOE result gave the mixing angle $(12\pm13)^\circ$ which is consistent with zero within the large error~\cite{Escribano:2007cd}. A phenomenological analysis of the processes $J/\psi(\psi')$ decaying into a pseudoscalar and a vector final states obtained the $\eta'$-glueball mixing angle to be around $9^\circ$ by considering the $\eta$-$\eta'$-glueball mixing model~\cite{Li:2007ky}. Obviously, the determined mixing angle varies in a fairly large range. Based on the $\eta'$-glueball mixing picture, there have been theoretical discussions on the possibility of $\eta(1405)$ as a pseudoscalar glueball candidate~\cite{Li:2007ky,Cheng:2008ss,He:2009sb,Li:2009rk,Tsai:2011dp}. However, the quenched lattice QCD studies~\cite{Morningstar:1997ff,Morningstar:1999rf,Chen:2005mg} predict that the mass of the pseudoscalar glueball is around 2.4–2.6 GeV, which is confirmed by lattice simulations with dynamical quarks~\cite{Chowdhury:2014mra,Richards:2010ck,Gregory:2012hu,Sun:2017ipk}. This raised a question on $\eta(1405)$ as a glueball candidate because of its much lighter mass. On the other hand, the strong hint for $\eta(1405)$ to be a glueball candidate is the observation that there exist three isoscalar pseudoscalar mesons $\eta(1295)$, $\eta(1405)$ and $\eta(1475)$ such that one of them is surplus according to the quark model. If $\eta(1405)$ and $\eta(1475)$   are the same state~\cite{Wu:2011yx}, there is no need for a pseudoscalar glueball state in this mass region. Some mixing model studies also favor the pseudoscalar glueball to have a mass heavier than 2 GeV~\cite{Mathieu:2009sg,Qin:2017qes}.

In this work, we will investigate the possible mixing of isoscalar pseudoscalar meson and the pseudoscalar glueball in $N_f=2$ lattice QCD. The isoscalar pseudoscalar meson is named $\eta$ throughout this work, which is the $\mathrm{SU(2)}$ counterpart of the $\mathrm{SU(3)}$ flavor singlet (approximately $\eta'$) in the $N_f=3$ case. We have generated a large ensemble of gauge configurations with $N_f=2$ degenerated $u,d$ quarks at a pion mass $m_\pi\approx 350$ MeV, so we can make a precise determination of $\eta$ mass. The calculation of $\eta'$ mass (and the $\eta'-\eta$ mixing) has been performed in several $N_f=2+1$ lattice QCD studies, whose results are in agreement with the physical value after the chiral extrapolation~\cite{Christ:2010dd,Michael:2013gka,Fukaya:2015ara,Ottnad:2017bjt}. A systematic and comprehensive lattice study on the properties of $\eta$ and $\eta'$ is presented in Ref.~\cite{Bali:2021qem}, where the masses, the decay constants and the gluonic matrix elements of $\eta$ and $\eta'$ have been derived to a high precision. There are also many studies on the $\eta$ mass from $N_f=2$ lattice QCD~\cite{CP-PACS:2002exu,Hashimoto:2008xg,Sun:2017ipk,Dimopoulos:2018xkm}. According to the Witten-Veneziano mechanism (WV), in the $N_f=2$ case, pion mass and $\eta$ mass are related as $m_\eta^2=m_\pi^2+m_0^2$, where $m_0^2$ is the correction from the topology induced interaction and is proportional to $N_f$. As a check of WV, we would like to take a look at this relationship and use the obtained $m_0^2$ to predict the $\eta'$ mass in the physical case (a pioneering work following this way in the quenched approximation can be found in Ref.~\cite{Kuramashi:1994aj}). After that, we calculate the correlation functions of the $\eta$ operator and the glueball operator, from which the mixing angle can be extracted. The strategy of the study is similar to that used in the $\eta_c$-glueball mixing~\cite{Zhang:2021xvl}. As an exploratory study, we tentatively treat the pseudoscalar glueball as a stable particle and ignore its resonance nature in the presence of light sea quarks. Obviously, the numeric task involves the calculation of the annihilation diagrams of $u,d$ quarks, so we adopt the distillation method~\cite{Peardon:2009gh} which enables the gauge covariant smearing of quark fields and the all-to-all quark propagators (perambulators) simultaneously. Since we also have the perambulators of the valence charm quark, we also calculate the $\eta$-$\eta_c$ correlation functions and explore their properties.

This paper is organized as follows: In Sec.~\ref{sec:numerical} we describe the lattice setup, operator construction and formulation of correlation functions. Section~\ref{sec:II} gives the theoretical formalism of the meson-glueball mixing, where the data analyses and the results can be found. The discussion and summary are given in Sec.~\ref{sec:summary}, and the preliminary results from $\eta$-$\eta_c$ correlation functions are presented here.

\section{Numerical Details}\label{sec:numerical}
\subsection{Lattice setup}\label{sec:setup}
\begin{figure}[t]
    \includegraphics[width=0.9\linewidth]{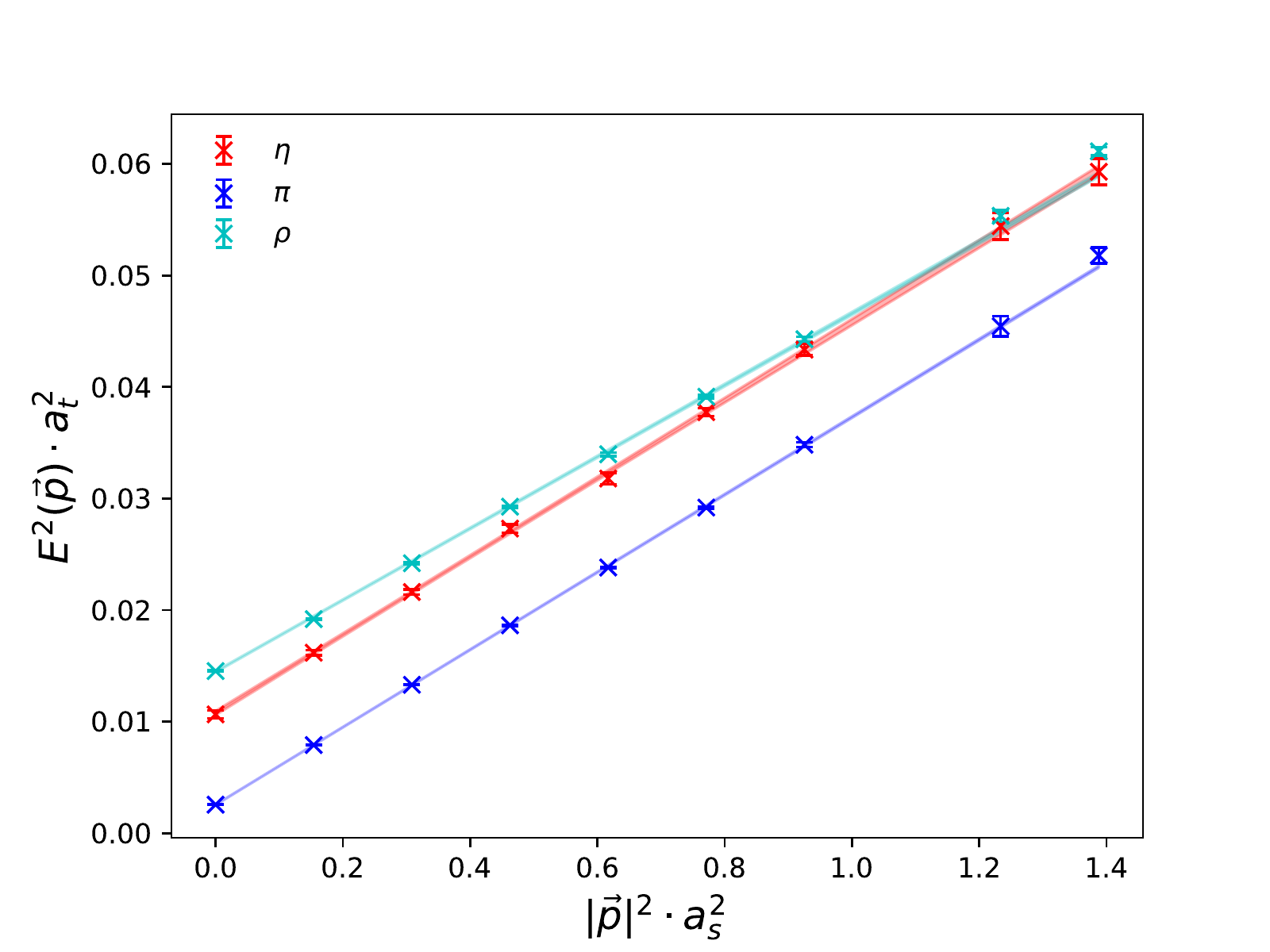}
    \caption{\label{fig:dispVPS} The dispersion relations of $\pi$ (blue) and $\eta$ (red). The data points are lattice results and the shaded lines illustrate the fitted results using Eq.~(\ref{eq:disp}). The best fit values of $\xi$ are 5.365(5) and 5.34(3) for $\pi$ and $\eta$, respectively. The cyan points are ground state energies obtained from the $\rho$ correlators at different spatial momentum $\vec{p}$, and the cyan line is the fitted result with $\xi=5.58(1)$ which deviates from $5.3$ drastically due to the interference with nearby $\pi\pi$ states.}
\end{figure}

We generate gauge configurations with $N_f=2$ degenerate $u,d$ on an $L^3\times T=16^3\times 128$ anisotropic lattice. We use the tadpole-improved Symanzik's gauge action for anisotropic lattices~\cite{Morningstar:1997ff,Chen:2005mg} and the tadpole-improved anisotropic clover fermion action~\cite{Edwards:2008ja,Sun:2017ipk}. The parameters in the action are tuned to give the anisotropy $\xi=a_s/a_t\approx 5.3$, where $a_t$ and $a_s$ are the temporal and spatial lattice spacings, respectively. The anisotropy $\xi\approx 5.3$ is checked by the dispersion relation of the pseudoscalar $\pi$ (along with that of $\eta$ calculated using the distillation method, see Sec.~\ref{sec:II}), which takes the continuum form
\begin{equation}~\label{eq:disp}
    E_X^2(\vec{p})a_t^2=m_X^2 a_t^2+\frac{1}{\xi^2} |\vec{p}|^2a_s^2,
\end{equation}
where $X$ refers to a specific hadron state and $\vec{p}$ is the spatial momentum $\vec{p}=\frac{2\pi}{La_s}\vec{n}$ on the lattice, and $\vec{n}$ is the mode of spatial momentum. Figure~\ref{fig:dispVPS} shows the energies obtained from the correlation functions of $\pi$, $\eta$ and $\rho$ at different spatial momentum modes up to $\vec{n}=(1,2,2)$. The data points of $\pi$ and $\eta$ fall on straight lines perfectly and can be well described by Eq.~(\ref{eq:disp}) with $\xi=5.365(5)$ and $5.34(3)$, respectively (illustrated as shaded lines in the figure). The fit to energies in the $\rho$ channel using Eq.~(\ref{eq:disp}) gives $\xi=5.58(1)$ which deviates from $5.3$ drastically. This can be definitely attributed to the influence of nearby $P$-wave $\pi\pi$ scattering states that have the same center-of-mass momentum as $\rho$. So we do not use $\rho$ meson to check the $\xi$ value.
%%%%%%%%%%%%%%%%%%%%%%%%%%%%%%%%%%%%%%%%%%%%%%%%%%%%%%%%%%%
%%%%%%%%%%%%%%%%%%%%%%%%%%%%%%%%%%%%%%%%%%%%%%%%%%%%%%%%%%%
%%%%%%%%%%%%%%%%%%%%%%%%%%%%%%%%%%%%%%%%%%%%%%%%%%%%%%%%%%%%%
\begin{table}[t]
    \centering\caption{\label{tab:v2-ps2} Experimental values of the masses of the pseudoscalar (P) and vector mesons (V) of quark configurations $n\bar{q}$, $n\bar{s}$, $n\bar{c}$, $s\bar{c}$, $n\bar{b}$ and $s\bar{b}$~\cite{Zyla:2020zbs}. Here $n$ refers to the light $u,d$ quarks. The right most column lists the $m_V^2-m_{PS}^2~(\mathrm{GeV}^2)$. In the row of $s\bar{s}$ states, the mass of the $s\bar{s}$ pseudoscalar $\eta_s$ is determined by the HPQCD collaboration from lattice QCD calculations~\cite{Davies:2009tsa}.}
    \begin{ruledtabular}
        \begin{tabular}{cccc}
            $q_l\bar{q}$ & $m_V$ (GeV) & $m_{PS}$ (GeV)              & $m_V^2-m_{PS}^2~(\mathrm{GeV}^2)$ \\
            \hline
            $n\bar{n}$   & 0.775       & 0.140                       & 0.581                             \\
            $n\bar{s}$   & 0.896       & 0.494                       & 0.559                             \\
            $s\bar{s}$   & 1.020       & 0.686~\cite{Davies:2009tsa} & 0.570                             \\
            $n\bar{c}$   & 2.010       & 1.870                       & 0.543                             \\
            $s\bar{c}$   & 2.112       & 1.968                       & 0.588                             \\
            $n\bar{b}$   & 5.325       & 5.279                       & 0.481                             \\
            $s\bar{b}$   & 5.415       & 5.367                       & 0.523                             \\                  
        \end{tabular}
    \end{ruledtabular}
\end{table}
%%%%%%%%%%%%%%%%%%%%%%%%%%%%%%%%%%%%%%%%%%%%%%%%%%%%%%%%%%%%

There are subtleties in the determination of the lattice spacing for our lattice setup. We intend to generate gauge configurations at a pion mass $m_\pi\sim 300~\text{MeV}$. As the first step, we calculate the static potential
\begin{equation}
    V(r)=V_0-\frac{e_c}{r}+\sigma r
\end{equation}
through Wilson loops to determine $e_c$ and the string tension $\sigma a_s^2$ (in lattice units). Then the spatial lattice spacing $a_s$ is determined in the unit of the Sommer's scale parameter $r_0$ according to the condition $\left.r^2\frac{dV}{dr}\right|_{r=r_0}=1.65$, namely
\begin{equation}~\label{eq:sommer}
    \frac{a_s}{r_0}=\sqrt{\frac{\sigma a_s^2}{1.65-e_c}}=0.334(2).
\end{equation}
We tentatively use the value $r_0=0.491(9)~\text{fm}$, which is determined at the physical point (in the chiral, continuum and infinite volume limits)~\cite{ETM:2015ned}, to set $a_s$ of our lattice, from which the quark mass is tuned to give $m_\pi\approx 300~\text{MeV}$. However, given this value of $a_s$, the mass of the vector meson $\rho$ is determined to be around 750 MeV, which is obviously lower than expected (note that $m_\rho$ is 770 MeV at the physical pion mass $m_\pi\sim 135$ MeV). This can be understood as follows. The measured value of $r_0/a_s$ actually also depends on the quark mass for a given lattice, such that an additional quark mass dependence is introduced to the values of $m_\pi$ and $m_\rho$ in terms of $r_0$. Since our quark mass is substantially far from the chiral limit, it is more reasonable to use the value of $r_0$ that is extrapolated from the physical pion mass to the comparable one in our situation. %may be attributed to the uncertainty of $r_0$ since its value varies from $0.45\sim 0.50$ fm determined by different lattice groups~\cite{FlavourLatticeAveragingGroup:2019iem}. 

An alternative scale setting scheme is to choose another quantity that is insensitive to quark masses. %Since hadron masses calculated on the lattice depend on both the quark mass parameter and the lattice spacing, the reasonable procedure is that one first sets the lattice spacing and then tunes the quark mass parameter to an expected value. So it is desirable to choose physical quantities that are insensitive to quark masses. 
Experimentally, there is an interesting relation between pseudoscalar meson masses $m_{PS}$ and the vector meson masses $m_V$ of the quark configuration $q_l\bar{q}$,
\begin{equation}\label{eq:dmsq}
    \Delta m^2 \equiv m_V^2 -m_{PS}^2\approx 0.56-0.58~~\mathrm{GeV}^2
\end{equation}
where $q_l$ stands for the $u,d,s$ quark and $q$ stands for $u,d,s,c$ quarks. The PDG results of the masses of these vector and pseudoscalar mesons ~\cite{Zyla:2020zbs} are collected in Table~\ref{tab:v2-ps2} along with their mass squared differences. Even though the reason is still unknown for this relation, empirically these values are insensitive to quark masses. On the other hand, the mass of $\eta_s$, the $s\bar{s}$ counterpart of $\pi$ (not considering the $s\bar{s}$ annihilation effects in the calculation), is determined to be $m_{\eta_s}=0.686(4)$ GeV from lattice QCD by the HPQCD collaboration~\cite{Davies:2009tsa}. Even though $\eta_s$ is not a physical state, the mass squared difference $m_\phi^2-m_{\eta_s}^2\approx 0.570 ~\mathrm{GeV}^2$ also satisfies the empirical relation in Eq.~(\ref{eq:dmsq}). In this study, the dimensionless masses of $\pi$ and $\rho$ is determined to be $m_\pi a_t=0.05055(13)$ and $m_\rho a_t=0.12046(20)$. In this sense, we assume the relation of Eq.~(\ref{eq:dmsq}) is somewhat general for heavy-light mesons and use it to set the scale parameter $a_t$. Of course, one should take caution to use this relation since $\rho$ is experimentally a wide resonance and decays into $P$-wave $\pi\pi$ states by 99\%. On our lattice the lowest $P$-wave $\pi\pi$ energy threshold in the rest frame of $\pi\pi$ is $2E_\pi(p)a_t\approx 0.1795$ with $\xi\approx 5.3$, which is substantially higher than $m_\rho a_t$. This means $\rho$ in its rest frame is a stable particle, such that the mass value $m_\rho$ is reliable. In practice, we make the least squares fitting to the mass squared differences over the $n\bar{n}$, $n\bar{s}$, $n\bar{c}$ and $s\bar{c}$ systems where $n$ refers to the $u,d$ quarks, and get the value $\overline{\Delta m^2}=0.568(8)$ $\mathrm{GeV}^2$, which serves as an input to give the lattice scale parameter $a_t^{-1}=6.894(51)$ GeV and the corresponding spatial lattice spacing $a_s\approx 0.1517(11)$ fm. We emphasize that the errors quoted in the values of $a_t$ and $a_s$ include only the statistical errors of $m_\pi a_t$ and $m_\rho a_t$, as well as the uncertainty of $\overline{\Delta m^2}$.
Accordingly, the $u,d$ mass parameter in this study gives $m_\pi=348.5(1.0)(2.6)$ MeV and $m_\rho=830.5(6.3)(6.1)$ MeV, where the second error is due to the uncertainty of $a_t^{-1}$. For most of this paper, we will not show the second error when listing our physical values. This means that most of the errors below are just statistical errors. We will bring the $a_t^{-1}$ error back to physical values in conclusion. Using the $m_\pi$ above, we have $m_\pi L_s \approx 3.9$ for this lattice setup, which warrants the small finite volume effects. The number of configurations of our gauge ensemble is 6991, which is crucial for the glueball-relevant studies. The details of the gauge ensemble are given in Table~\ref{tab:config}.
\begin{table}[t]
    \renewcommand\arraystretch{1.5}
    \caption{Parameters of the gauge ensemble.}
    \label{tab:config}
    \begin{ruledtabular}
        \begin{tabular}{lllllc}
            $L^3 \times T$    & $\beta$ & $a_t^{-1}$(GeV) & $\xi$      & $m_\pi$(MeV) & $N_\mathrm{cfg}$ \\\hline
            $16^3 \times 128$ & 2.0     & $6.894(51)$     & $\sim 5.3$ & $348.5(1.0)$ & $6991$           \\
        \end{tabular}
    \end{ruledtabular}
\end{table}
%If the determined $a_s$ is put in to Eq.~(\ref{eq:sommer}), the $r_0$ at our $m_\pi\approx 350~\text{MeV}$ is estimated to be $r_0\approx 0.455(3)~\text{fm}$, whose deviation from $r_0^\text{phys}=0.491(9)~\text{fm}$ manifests the $m_\pi$ dependence of $r_0$ to some extent.
%As a cross-check of this scheme, we also calculate the heavy quark static potential through Wilson loops, from which the relation of the Sommer's scale parameter $r_0$ to the spatial lattice spacing $a_s$ is expressed as $\frac{r_0}{a_s}=\sqrt{\frac{1.65-e_c}{\sigma a_s^2}}$, where $e_c$ and $\sigma a_s^2$ are the parameters of the Cornell type parametrization of the static potential. Using the obtained $a_t$ and $\xi$, we estimate the $r_0$ to be $0.455(3)$ fm.

%%%%%%%%%%%%%%%%%%%%%%%%%%%%%%%%%%%%%%%%%%%%%%%%%%%%%%%%%%%
\begin{figure}[t]
    \includegraphics[width=0.9\linewidth]{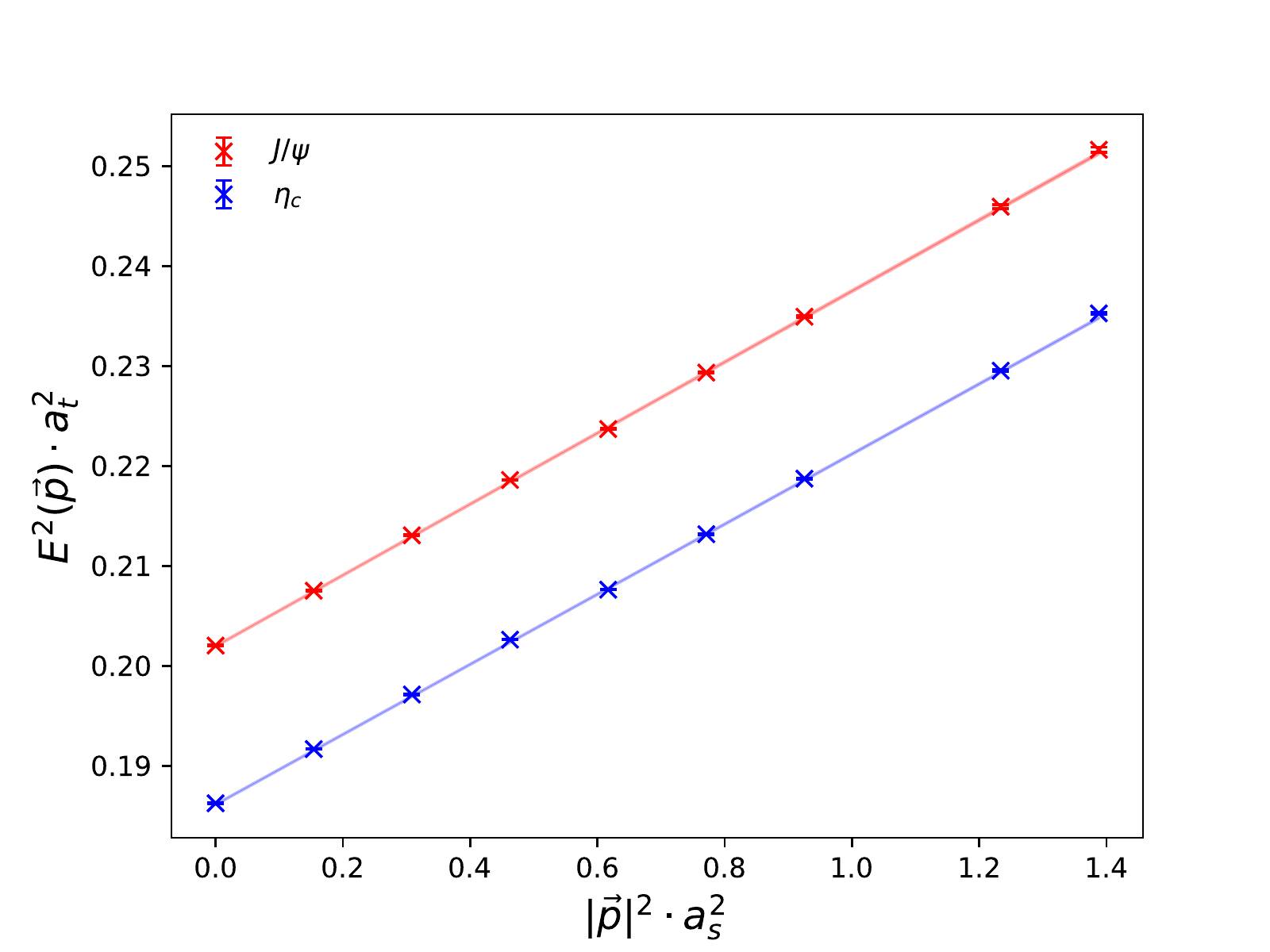}
    \caption{\label{fig:dispcharm}The dispersion relations of $\eta_c$ and $J/\psi$. The data points are lattice results and the shaded lines illustrate the fit results using Eq.~(\ref{eq:disp}). The fitted results of $\xi$ are 5.341(2), 5.307(5) for $\eta_c$ and $J/\psi$, respectively.}
\end{figure}
%%%%%%%%%%%%%%%%%%%%%%%%%%%%%%%%%%%%%%%%%%%%%%%%%%%%%%%%%%%
For the valence charm quark, we adopt the clover fermion action in Ref.~\cite{CLQCD:2009nvn}, and the charm quark mass parameter is tuned to give $(m_{\eta_c}+3m_{J/\psi})/4=3.069$ GeV. With the tuned charm quark mass parameter, we generate the perambulators of charm quark on our ensemble, from which the masses of $\eta_c$ and $J/\psi$ are derived precisely to be $m_{\eta_c}=2.9750(3)$ GeV and $m_{J/\psi}=3.0988(4)$ GeV. The according $1S$ hyperfine splitting is $\Delta_\mathrm{HFS}=m_{J/\psi}-m_{\eta_c}=123.8(5)$ MeV. We also check the dispersion relation in Eq.~(\ref{eq:disp}) for $\eta_c$ and $J/\psi$ up to the momentum mode $\vec{n}=(1,2,2)$. As shown in Fig.~\ref{fig:dispcharm}, the dispersion relation is almost perfectly satisfied with $\xi=5.341(2)$ and $5.307(5)$ for $\eta_c$ and $J/\psi$, respectively.

As a further check of our scale setting scheme, we also calculate the masses of $D$ and $D^*$ and get $m_D=1.882(1)$ GeV and $m_D^*=2.023(1)$ GeV on a fraction of configurations of our ensemble. The hyperfine splitting $\Delta_\mathrm{HFS}(D)=m_{D^*}-m_D=0.141(2)$ GeV almost reproduces the experimental values $m_{D^{*0}}-m_{D^0}=0.14201(7)$ GeV and $m_{D^{*+}}-m_{D^+}=0.14060(7)$ GeV~\cite{Zyla:2020zbs}. This manifests that our tuning of charm quark mass works well. The dispersion relation Eq.~(\ref{eq:disp}) is also checked to be correct for $D$ and $D^*$ with $\xi=5.32(2)$ and $5.31(3)$, respectively. The figure is similar to Fig.~\ref{fig:dispVPS} and Fig.~\ref{fig:dispcharm} and is omitted here to save space.

Table~\ref{tab:charm} collects the results of $\eta_c$, $J/\psi$, $D$ and $D^*$ mesons. Along with the values of $\xi$ derived from the dispersion relations of $\pi$ and $\eta$, we can see that the values of $\xi$ for different mesons are in agreement with the value $\xi=5.30$ during the parameter tuning and are consistent with each other within 1\%.

\begin{table}[t]
    \renewcommand\arraystretch{1.5}
    \caption{The masses of $J/\psi$, $\eta_c$, $D$ and $D^*$. The PDG value of $m_{D^{(*)}}$ is the average of masses of $D^{(*)0}$ and $D^{(*)+}$.}
    \label{tab:charm}
    \begin{ruledtabular}
        \begin{tabular}{llllc}
            $X$                     & $\eta_c$  & $J/\psi$  & $D$          & ${D^*}$      \\\hline
            $m_X$(GeV)              & 2.9750(3) & 3.0988(4) & 1.882(1)     & 2.023(1)     \\
            PDG~\cite{Zyla:2020zbs} & 2.983     & 3.097     & $\sim 1.867$ & $\sim 2.008$ \\
            $\xi$                   & 5.341(2)  & 5.307(5)  & 5.32(2)      & 5.31(3)
        \end{tabular}
    \end{ruledtabular}
\end{table}

\subsection{Operator construction and distillation method}
The principal goal of this work is to investigate the possible mixing of the pseudoscalar glueball and the pseudoscalar $q\bar{q}$ meson, therefore the quark annihilation diagrams should be taken care of. For this to be done, we adopt the distillation method~\cite{Peardon:2009gh} which provides a smearing scheme for quark fields (Laplacian Heaviside smearing) and the calculation strategy of the all-to-all propagators of the smeared quark fields that are distilled to be perambulators in the Laplacian Heaviside subspace of the spatial Laplacian operator $-\nabla^2$. Since we plan to investigate the $\eta-\eta_c$ correlation functions as well, we calculate the perambulators of $u,d$ and $c$ quarks on our large gauge ensemble in the Laplacian Heaviside space spanned by the $N=70$ eigenvectors of $-\nabla^2$ operator with the lowest eigenvalues. For the pseudoscalar glueball operator, we adopt the strategy in Refs.~\cite{Morningstar:1999rf,Chen:2005mg} to get the optimized Hermitian operator $\mathcal{O}_G(t)=\mathcal{O}^\dagger_G(t)$ coupling mainly to the ground state glueball based on different prototypes of Wilson loops and gauge link smearing schemes (see Appendix for the details).

For the isoscalar $\eta$, the interpolation field can be defined as
\begin{equation}
    \mathcal{O}_\Gamma = \frac{1}{\sqrt{2}}\left[\bar{u}^{(s)}\Gamma u^{(s)} + \bar{d}^{(s)}\Gamma d^{(s)}\right],
\end{equation}
where $\Gamma$ refers to $\gamma_5$ or $\gamma_4\gamma_5$, $u^{(s)}$ and $d^{(s)}$ are Laplacian Heaviside smeared $u,d$ quark fields. Thus the correlation function of $\mathcal{O}_\Gamma$ can be expressed as
\begin{eqnarray}\label{eq:ccc}
    C_{\Gamma\Gamma}(t)&=&\frac{1}{T}\sum\limits_{t_s=1}^{T}\sum\limits_{\mathbf{xy}}\langle \mathcal{O}_\Gamma (\mathbf{x},t+t_s)\mathcal{O}_\Gamma ^\dagger(\mathbf{y},t_s)\rangle\nonumber\\
    &\equiv& \mathcal{C}_\Gamma(t)+2\mathcal{D}_\Gamma(t)
\end{eqnarray}
with $\mathcal{C}_\Gamma(t)$ and $\mathcal{D}_\Gamma(t)$ being the contributions from the connected and disconnected diagrams, respectively. We also consider the following correlation functions
\begin{eqnarray}\label{eq:corrs}
    C_{GG}(t)&=&\frac{1}{T}\sum\limits_{t_s=1}^{T}\langle \mathcal{O}_G(t+t_s)\mathcal{O}_G(t_s)\rangle\nonumber\\
    C_{G\Gamma}(t)&=&\frac{1}{T}\sum\limits_{t_s=1}^{T}\sum\limits_{\mathbf{x}}\langle \mathcal{O}_G(t+t_s)\mathcal{O}_\Gamma^\dagger(\mathbf{x},t_s)\rangle\nonumber\\
    C_{\Gamma G}(t)&=&\frac{1}{T}\sum\limits_{t_s=1}^{T}\sum\limits_{\mathbf{x}}\langle \mathcal{O}_\Gamma(\mathbf{x},t+t_s)\mathcal{O}_G(t_s)\rangle\nonumber\\
    &=&\mp C_{G\Gamma}(t)\nonumber\\
    C_{\Gamma\Gamma_c}(t)&=&\frac{1}{T}\sum\limits_{t_s=1}^{T}\sum\limits_{\mathbf{xy}}\langle \mathcal{O}_{\Gamma}(\mathbf{x},t+t_s)\mathcal{O}_{\Gamma_c}^\dagger(\mathbf{y},t_s)\rangle
\end{eqnarray}
where the $\mp$ sign comes from the Hermiticity of $\mathcal{O}_{\Gamma}$, it takes minus sign for $\Gamma=\gamma_5$ (anti-Hermitian) and plus sign for $\gamma_4\gamma_5$ (Hermitian). The operator $\mathcal{O}_{\Gamma_c}=\bar{c}^{(s)}\Gamma_c c^{(s)}$ where $\Gamma_c=\gamma_5,\gamma_4\gamma_5$ is also defined in terms of the Laplacian Heaviside smeared charm quark field $c^{(s)}$. Obviously, all of these correlation functions except for $C_{GG}(t)$ are contributed by quark annihilation diagrams and can be dealt with conveniently in the framework of the distillation method.

\subsection{$\eta$ mass as a further calibration}\label{sec:etamass}
%%%%%%%%%%%%%%%%%%%%%%%%%%%%%%%%%%%%%%%%%%%%%%%%%%%%%%%%%%%
\begin{figure}[t]
    \includegraphics[width=0.9\linewidth]{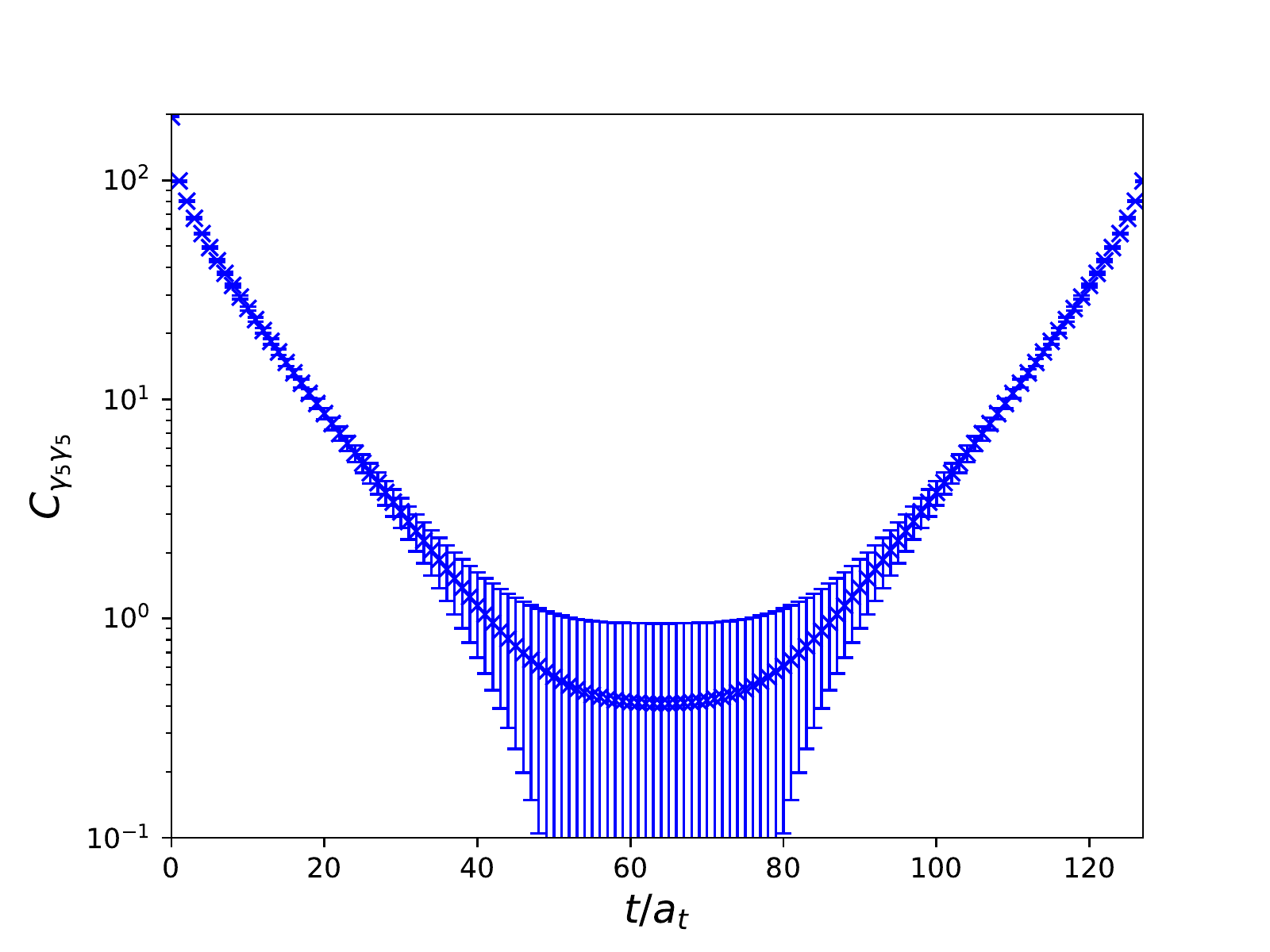}
    \caption{\label{fig:corr-gamma5}The correlation function of $\eta$ with $\Gamma=\gamma_5$. We can see the function contains a nonzero constant with a large error near $t=T/2$.}
\end{figure}
\begin{figure}[t]
    \includegraphics[width=0.9\linewidth]{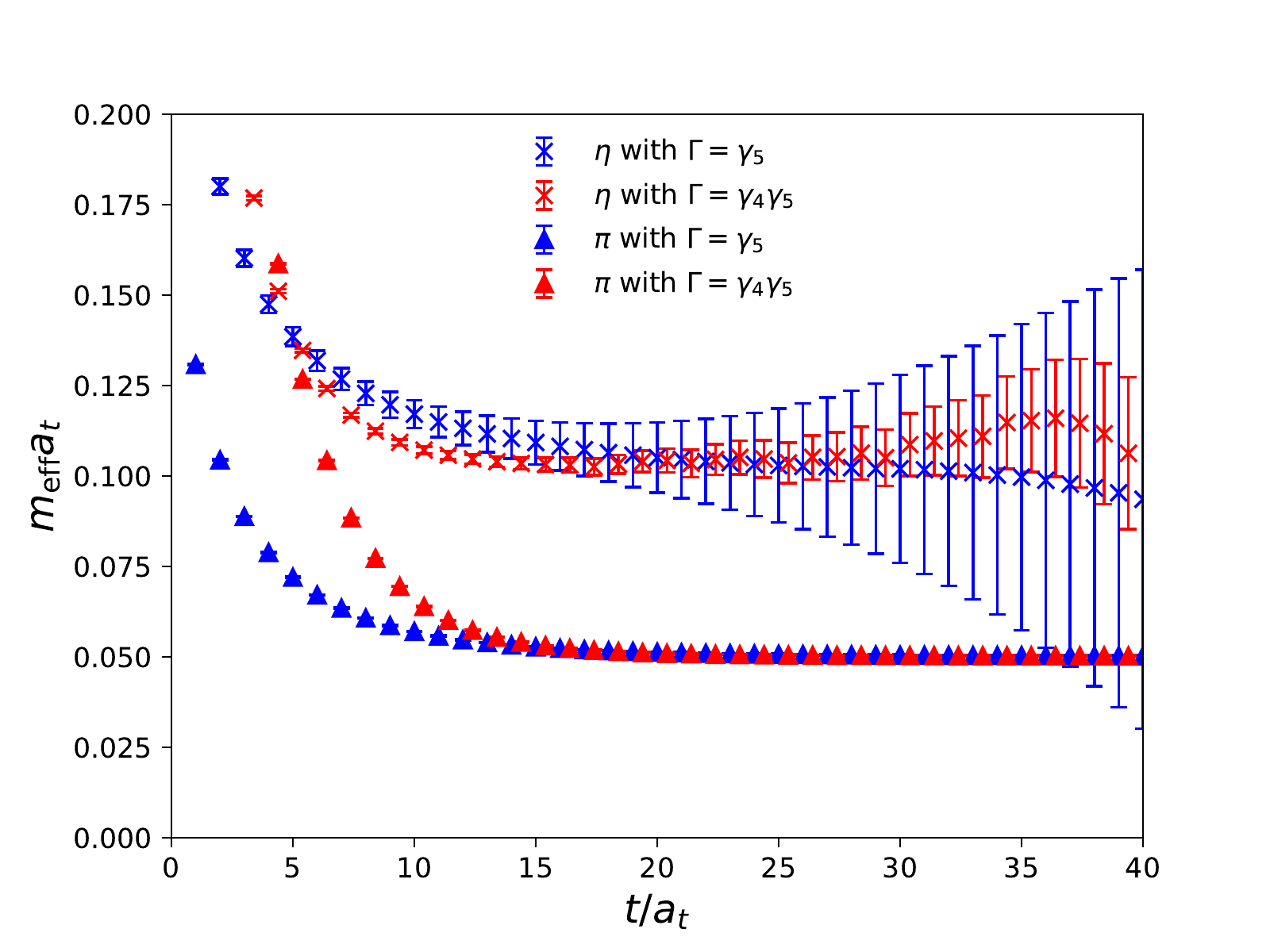}
    \caption{\label{fig:effm-5}The effective masses of $\pi$ and $\eta$ with different $\Gamma$ insertions. For $\eta$, the signal-to-noise ratio looks better in the case $\Gamma=\gamma_4\gamma_5$ because the corresponding correlation function doesn't contain a constant associated with the topology of QCD vacuum.}
\end{figure}

%%%%%%%%%%%%%%%%%%%%%%%%%%%%%%%%%%%%%%%%%%%%%%%%%%%%%%%%%%%
We calculate two types of correlation functions for $\eta$, namely $C_{\gamma_5\gamma_5}(t)$ and $C_{(\gamma_4\gamma_5)(\gamma_4\gamma_5)}(t)$. We do observe the finite volume artifact that $C_{\gamma_5\gamma_5}(t)$ approaches to a nonzero constant when $t$ is large, as shown in Fig.~\ref{fig:corr-gamma5}. It has been argued that this constant term comes from the topology of QCD vacuum and can be approximately expressed as $a^5(\chi_\mathrm{top} + Q^2/V)/T$ where $a$ is the lattice spacing (in the isotropic case), $\chi_\mathrm{top}$ is the topological susceptibility, $Q$ is the topological charge, $V$ is the spatial volume and $T$ is the temporal extension of the lattice~\cite{Aoki:2007ka,Bali:2014pva,Dimopoulos:2018xkm}. This can be understood from the anomalous axial vector current relation that the $\mathcal{O}_{\gamma_5}$ has a direct connection with the topological charge density operator.
It is interesting to observe that $C_{(\gamma_4\gamma_5)(\gamma_4\gamma_5)}(t)$ has normal large $t$ behavior that it damps to zero for large $t$. As usual, the $t$-behavior of these correlation functions can be seen more clearly through their effective mass functions defined as
\begin{equation}\label{eq:geffm}
    m_\mathrm{eff}(t)=\ln \frac{C_{\Gamma\Gamma}(t)}{C_{\Gamma\Gamma}(t+1)}
\end{equation}
where $\Gamma$ refers to $\gamma_5$ or $\gamma_4\gamma_5$. Figure~\ref{fig:effm-5} shows these effective mass functions. Benefited from the large statistics of our gauge ensemble, the effective mass plateau starts from $t\sim 10$, and the signal-to-noise ratio keeps good for $t$ beyond 20 in the case $\Gamma=\gamma_4\gamma_5$.
%%%%%%%%%%%%%%%%%%%%%%%%%%%%%%%%%%%%%%%%%%%%%%%%%%%%%%%%%%%
%%%%%%%%%%%%%%%%%%%%%%%%%%%%%%%%%%%%%%%%%%%%%%%%%%%%%%%%%%%
By a combined fit of $C_{\gamma_5\gamma_5}(t)$ and $C_{(\gamma_4\gamma_5)(\gamma_4\gamma_5)}(t)$ using two mass terms, we obtain the best-fit mass of $\eta$ to be
\begin{equation}\label{eq:meta}
    m_\eta=714.1(5.8)~~\mathrm{MeV}.
\end{equation}
where the error is obtained through the jackknife analysis. Considering the uncontrolled systematic uncertainties in our calculation (the continuum extrapolation and the chiral extrapolation have not been performed), this value is compatible with other determinations~\cite{CP-PACS:2002exu,Hashimoto:2008xg,Dimopoulos:2018xkm}. According to WV, the large mass of the flavor singlet pseudoscalar meson has a direct connection with the topology of QCD vacuum. In the $N_f=2$ case, to the leading order of chiral perturbation theory, $m_\eta$ is related to $m_\pi$
\begin{equation}
    m_\eta^2 = m_\pi^2 +m_0^2,
\end{equation}
where the parameter $m_0^2$ is defined in terms of the topological susceptibility $\chi_\mathrm{top}$ and the pion decay constant $f_\pi$, namely
\begin{equation}
    m_0^2 = m_\eta^2-m_\pi^2\approx\frac{4N_f}{f_\pi^2} \chi_\mathrm{top}.
\end{equation}
Usually, $\chi_\mathrm{top}$ is for the pure gauge case and is expected to be independent of flavor number $N_f$. Using the values of $m_\pi=348.5(1.0)$ MeV and $m_\eta$ in Eq.~(\ref{eq:meta}), $m_0^2$ can be derived to be $m_0^2=0.3885(77)$ $\mathrm{GeV}^2$. According to the leading order chiral perturbation theory for $N_f=2$, the $m_\pi$ dependence of $f_\pi$ is~\cite{Colangelo:2001df} 
\begin{equation}
    f_\pi=\sqrt{2}F \left[1+\xi \left(\ln \frac{m_\pi^{\mathrm{phys},2}}{m_\pi^2}+l_4\right)+\mathcal{O}(\xi^2)\right]
\end{equation}
where $m_\pi^\mathrm{phys}=134.98$ MeV is the physical pion mass, $\xi$ is approximated by $m_\pi^2/(4\pi F)^2$, the pion decay constant in the chiral limit $F = 86(1)$ MeV and the low energy constant $l_4=4.40(28)$ are taken from FLAG 2019~\cite{FlavourLatticeAveragingGroup:2019iem}. 
For our $m_\pi=348.5(1.0)$ MeV, we have 
\begin{equation}
    \frac{f_\pi}{f_\pi^\mathrm{phys}}=1.17(1)_{-0.02}^{+0.03}
\end{equation}
where the first error comes from the error of $F$ and the second one is due to the uncertainty of $l_4$. Using this value and $f_\pi^\text{phys}=130.2(8)~\text{MeV}$~\cite{RBC:2014ntl,Follana:2007uv,MILC:2010hzw,FlavourLatticeAveragingGroup:2019iem}, $\chi_\mathrm{top}^{1/4}$ is estimated to be $\chi_\mathrm{top}^{1/4}= 183(3)$ MeV, which is very close to the phenomenological value $180$ MeV and the lattice value $185.3(5.6)$ MeV~\cite{Cichy:2015jra}. In the physical $N_f=3$ case at physical pion mass, the GMOR relation implies the mass of the singlet counterpart of the pseudoscalar octet should be $m_1^2=(2m_K^2 + m_\pi^2)/3=0.170~\mathrm{GeV}^2$. Thus using the topological susceptibility we obtained, we can estimate the mass of the flavor singlet pseudoscalar meson to be
\begin{equation}
    m_{\eta_1}^2=m_1^2 +\frac{12 \chi_\mathrm{top}}{f_\pi^2}\approx 0.961~\mathrm{GeV}^2,
\end{equation}
which corresponds to $m_{\eta_1}\approx 0.981(29)$ GeV and is not far from the experimental value $m_{\eta'} =0.958~{\mathrm{GeV}}$. The GMOR relation also implies $m_{\eta_8}^2=(4m_K^2-m_\pi^2)/3\approx 0.321~\mathrm{GeV}^2$.
%One can test that the relation $m_{\eta_1}^2 +m_{\eta_8}^2 = m_\eta^{\mathrm{exp.},2} +m_{\eta'}^2$ is satisfied within 2\%. 
This confirms again that the Witten-Veneziano mechanism for the flavor singlet pseudoscalar mass works fairly well both for the 
$\mathrm{SU(2)}$ and $\mathrm{SU(3)}$ flavor symmetry. 

%On the other hand, these results also reinforce the reasonability of our scale setting in Sec.~\ref{sec:numerical}.

\section{Toward the $\eta$-glueball mixing}\label{sec:II}
\subsection{Theoretical consideration}\label{sec:formalism}

In order to investigate the possible mixing between the pseudoscalar glueball and $\eta$, we must parametrize the correlation functions in Eq.~(\ref{eq:corrs}). We adopt the following theoretical logic. As usual in the lattice study, the correlation function $C_{XY}(t)$ of operator $\mathcal{O}_X$ and $\mathcal{O}_Y$ can be parametrized as
\begin{equation}\label{eq:gc}
    C_{XY}\approx \sum\limits_{n\neq 0} \left[ \langle 0|\mathcal{O}_X|n\rangle \langle n|\mathcal{O}^\dagger_{Y}|0\rangle \left(e^{-E_nt}\pm e^{-E_n(T-t)}\right)\right]
\end{equation}
where the $\pm$ sign is for the same and opposite Hermiticities of $\mathcal{O}_X$ and $\mathcal{O}_Y$, respectively, and $|n\rangle$ are the eigenstates of the lattice Hamiltonian $\hat{H}$ defined as $\hat{H}|n\rangle=E_n|n\rangle$ with $E_n$ being the corresponding eigen energies. For a given quantum number, $|n\rangle$'s establish an orthogonal and complete set, namely, $\sum\limits_n |n\rangle\langle n|=1$ with the normalization condition $\langle m|n\rangle=\delta_{mn}$. In principle, $\hat{H}$ only exists heuristically, so we do not know the exact particle configurations of these eigenstates simply from the correlation function in Eq.~(\ref{eq:gc}). As far as the flavor singlet pseudoscalar channel is concerned in the $N_f=2$ QCD theory, each of the state $|n\rangle$ should be a specific admixture of bare $\eta$ states and bare glueballs if they exist theoretically (here we ignore the multihadron states temporarily) and can be taken as the states in the eigenstate set $\{|\alpha_n\rangle,n=1,2,\ldots\}$ of the free Hamiltonian $\hat{H}_0$. Since we are working in a unitary lattice framework for $u,d$ quarks, in principle this state set is orthogonal and complete with the normalization condition $\langle \alpha_m|\alpha_n\rangle=\delta_{mn}$. Now we introduce the interaction Hamiltonian $\hat{H}_I$ to account for the dynamics of the possible mixing, such that $|n\rangle $ of $\hat{H}=\hat{H}_0+\hat{H}_I$ can be expanded in terms of $|\alpha_m\rangle$ as
\begin{equation}
    |n\rangle =\sum\limits_m C_{nm}|\alpha_m\rangle
\end{equation}
with $\sum\limits_m |C_{nm}|^2=1$. In this sense, one can say that $|n\rangle$ is an admixture of states $|\alpha_m\rangle$ whose fractions are $|C_{nm}|^2$, respectively. Furthermore, if $\hat{H}_I$ is small relative to $\hat{H}_0$, then to the lowest order of the perturbation theory, one has
\begin{eqnarray}
    |n\rangle&=&|\alpha_n\rangle+\sum\limits_{m\ne n}\frac{\langle \alpha_m|\hat{H}_I|\alpha_n\rangle}{E_n^{(0)}-E_m^{(0)}}|\alpha_m\rangle\nonumber\\
    E_n&=& E_n^{(0)}+\sum\limits_{m\ne n}\frac{|\langle \alpha_m|\hat{H}_I|\alpha_n\rangle|^2}{E_n^{(0)}-E_m^{(0)}}
\end{eqnarray}
where $E^{(0)}_n$ is the eigenenergy of $|\alpha_n\rangle$ and is ordered from low to high.

The experimentally observed isoscalar pseudoscalars are $\eta$, $\eta'$, $\eta(1295)$, $\eta(1405/1475)$, etc.~\cite{Zyla:2020zbs}, which are identified as $I=0$ members of different $\bar{q}q$ $\mathrm{SU(3)}$ nonets of different radial quantum numbers. As for the $\mathrm{SU(2)}$ case of this study, since there is only one isoscalar in each isospin quartet, the spectrum can be simplified largely. Because the smeared quark field suppresses the contribution of excited states in the correlation functions of $\eta$, as a simple approximation, we can truncate the spectrum of $\eta$ state to be $\eta$ and $\eta^*$ with $\eta^*$ taking account into all the excited $\eta$ states. On the other hand, quenched lattice QCD predicted the mass of the lowest pseudoscalar glueball is around 2.4–2.6 GeV. This seems to be confirmed by the correlation function $C_{GG}(t)$ of the optimized operator $\mathcal{O}_G$, which is expected to couple most to the ground state. So we include the ground state pseudoscalar glueball $|G\rangle$ and another state $|G^*\rangle $ in the state basis $\{|\alpha_i\rangle, i=1,2,\cdots\}$ with $|G^*\rangle$ standing for all the excited states of the pseudoscalar glueball. Finally, we have the following state basis
\begin{equation}
    |\alpha_i\rangle=|\eta\rangle, |\eta^*\rangle, \ldots, |G\rangle, |G^*\rangle,\ldots,
\end{equation}
With this state basis, the free Hamiltonian $\hat{H}_0=\mathrm{diag}\{m_\eta, m_{\eta^*},m_G, m_{G^*}\}$ is diagonal with the matrix elements being the bare masses of the basis states, respectively, and being ordered from low to high. Theoretically, $|\eta\rangle$ and $|\eta^*\rangle$ are orthogonal, so do the states $|G\rangle$ and $|G^*\rangle$. Thus the interaction Hamiltonian $\hat{H}_I$ can be expressed as
\begin{equation}
    H_I=\left(
    \begin{array}{cccc}
        0   & 0   & x_1 & y_1 \\
        0   & 0   & x_2 & y_2 \\
        x_1 & x_2 & 0   & 0   \\
        y_1 & y_2 & 0   & 0   \\
    \end{array}
    \right),
\end{equation}
where $x_i, y_i$ are called mixing energies sometimes. Accordingly, we have the following state expansion of $|n\rangle$
\begin{eqnarray}\label{eq:expansion}
    |1\rangle&\approx&|\eta\rangle +\frac{x_1}{m_\eta-m_G}|G\rangle+\frac{y_1}{m_\eta-m_{G^*}}|G^*\rangle\nonumber\\
    |2\rangle&\approx&|\eta^*\rangle +\frac{x_2}{m_{\eta^*}-m_G}|G\rangle+\frac{y_2}{m_{\eta^*}-m_{G^*}}|G^*\rangle\nonumber\\
    |3\rangle&\approx&|G\rangle +\frac{x_1}{m_G-m_\eta}|\eta\rangle+\frac{x_2}{m_G-m_{\eta^*}}|\eta^*\rangle\nonumber\\
    |4\rangle&\approx&|G^*\rangle +\frac{y_1}{m_{G^*}-m_\eta}|\eta\rangle+\frac{ y_2}{m_{G^*}-m_{\eta^*}}|\eta^*\rangle.
\end{eqnarray}

%%%%%%%%%%%%%%%%%%%%%%%%%%%%%%%%%%%%%%%%%%%%%%%%%%%%%%%%%%%%%%%%%%%%%%%%%%
\begin{figure}[t]
    \includegraphics[width=0.9\linewidth]{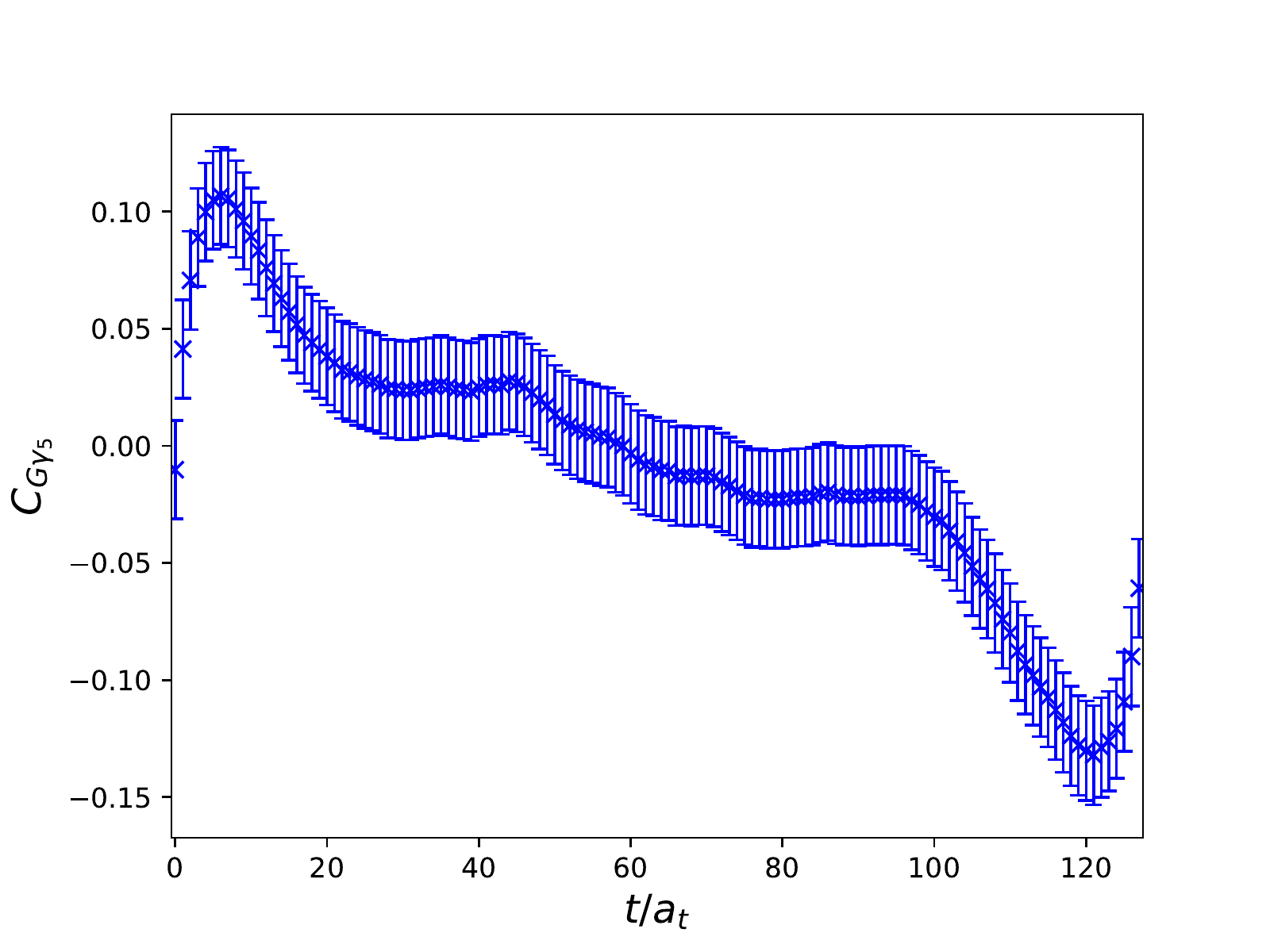}
    \caption{\label{fig:corr-eta-G-origin}The correlation function of $\eta-G$ with $\Gamma=\gamma_5$, called $C_{G\gamma_5}$. The value of this correlator tends to zero at $t=0$ within errors, and the error seems to be a constant independent of $t$. The correlator approaches a positive(negative) constant when $t<\frac{T}{2}$($t>\frac{T}{2}$) which might be due to the contribution from topology.}
\end{figure}
%%%%%%%%%%%%%%%%%%%%%%%%%%%%%%%%%%%%%%%%%%%%%%%%%%%%%%%%%%%%%%%%%%%%%%%%%%
\subsection{The $\Gamma=\gamma_5$ case}
Now we explore the possibility of glueball-$\eta$ mixing through the correlation function $C_{G\gamma_5}(t)$. Let us take a look at the $t$-dependence
of $C_{G\gamma_5}(t)$ shown in Fig.~\ref{fig:corr-eta-G-origin}. We have the following observations:
\begin{itemize}
    \item[] (1)~ $C_{G\gamma_5}(t)$ is anti-symmetric with respect to $t=T/2$ and tends to 0 at $t=0$. This is understood because $\mathcal{O}_G$ is hermitian by construction (and even under the time reversal transformation $\mathcal{T}$) while $\mathcal{O}_{\gamma_5}$ is anti-Hermitian and $\mathcal{T}$-odd. At $t=0$, since the product $\mathcal{O}_G(0)\mathcal{O}_{\gamma_5}(0)$ is $\mathcal{T}$-odd, its vacuum expectation value $\langle \mathcal{O}_G\mathcal{O}_{\gamma_5}(0)\rangle$ certainly vanishes.
    \item[] (2)~ $C_{G\gamma_5}(t)$ approaches a positive (negative) constant when $t<\frac{T}{2}$ ($t>\frac{T}{2}$). This may be due to the constant contribution from the topology similar to the case of $C_{\gamma_5\gamma_5}(t)$ discussed in Sec.~\ref{sec:etamass}. Since $C_{G\gamma_5}(t)$ is now $\mathcal{T}$-odd, so does the topology contribution.
    \item[] (3)~ Even though its central value is smooth, the error of $C_{G\gamma_5}(t)$ is almost constant throughout the time range.
\end{itemize}
In order to find a function form to describe the time behavior of $C_{G\gamma_5}(t)$, we take the following approximations
\begin{eqnarray}\label{eq:create}
    \mathcal{O}_G^\dagger|0\rangle&\approx&\sum\limits_{i\neq 0}\sqrt{Z_{G_i}}|G_i\rangle\nonumber\\
    \mathcal{O}_{\gamma_5}^\dagger|0\rangle&\approx&\sum\limits_{i\neq 0}\sqrt{Z_{\gamma_5,i}}|\eta_i\rangle,
\end{eqnarray}
by the assumptions $\langle 0|\mathcal{O}_G|\eta_i\rangle\approx 0$ and $\langle 0|\mathcal{O}_{\gamma_5}|G_i\rangle\approx 0$ similar to those adopted in the $\eta-\eta'$ mixing studies~\cite{Christ:2010dd,Michael:2013gka}.

Consequently, we have the following coupling matrix elements of the operators $\mathcal{O}_G$ and $\mathcal{O}_{\gamma_5}$,
\begin{eqnarray}\label{eq:gcouple}
    \langle 0|\mathcal{O}_G|1\rangle&=&\frac{x_1 \sqrt{Z_G}}{m_\eta-m_G}+\frac{y_1 \sqrt{Z_{G^*}}}{m_\eta-m_{G^*}}\nonumber\\
    \langle 0|\mathcal{O}_G|2\rangle&=&\frac{x_2 \sqrt{Z_G}}{m_{\eta^*}-m_G}+\frac{y_2 \sqrt{Z_{G^*}}}{m_{\eta^*}-m_{G^*}}\nonumber\\
    \langle 0|\mathcal{O}_G|3\rangle&=&\sqrt{Z_G}\nonumber\\
    \langle 0|\mathcal{O}_G|4\rangle&=&\sqrt{Z_{G^*}}
\end{eqnarray}
and
\begin{eqnarray}\label{eq:etacouple}
    \langle 0|\mathcal{O}_{\gamma_5}|1\rangle&=&\sqrt{Z_{\gamma_5,1}}\nonumber\\
    \langle 0|\mathcal{O}_{\gamma_5}|2\rangle&=&\sqrt{Z_{\gamma_5,2}}\nonumber\\
    \langle 0|\mathcal{O}_{\gamma_5}|3\rangle&=&-\frac{x_1 \sqrt{Z_{\gamma_5,1}}}{m_\eta-m_G}-\frac{x_2 \sqrt{Z_{\gamma_5,2}}}{m_{\eta^*}-m_{G}}\nonumber\\
    \langle 0|\mathcal{O}_{\gamma_5}|4\rangle&=&-\frac{y_1 \sqrt{Z_{\gamma_5,1}}}{m_{\eta}-m_{G^*}}-\frac{y_2 \sqrt{Z_{\gamma_5,2}}}{m_{\eta^*}-m_{G^*}}
\end{eqnarray}

As an exploratory study and for the simplicity of the future data analysis, we temporarily neglect $\eta^*$ contributes to $C_{G\gamma_5}(t)$. That is to say, we take the further approximation $\mathcal{O}_{\gamma_5}^\dagger|0\rangle\approx \sqrt{Z_{\gamma_5,1}}|\eta\rangle$. Thus after inserting Eqs.~(\ref{eq:expansion}), (\ref{eq:create}), (\ref{eq:gcouple}), (\ref{eq:etacouple}) to Eq.~(\ref{eq:gc}) and ignore the terms relevant to $|\eta^*\rangle$, we have the approximate expression of $C_{G\gamma_5}(t)$ as
\begin{eqnarray}\label{eq:gc_general}
    C_{G\gamma_5}(t)&\approx&\sqrt{Z_G Z_{\gamma_5,1}}\frac{x_1 }{m_\eta-m_G} \left(e^{-m_1 t}-e^{-m_3 t}\right)\nonumber\\
    &+&\sqrt{Z_{G^*} Z_{\gamma_5,1}}\frac{y_1 }{m_{\eta}-m_{G^*}} \left(e^{-m_1 t}-e^{-m_4 t}\right)\nonumber\\
    &-& (t\to (T-t)~~ \mathrm{terms}).
\end{eqnarray}
The feature of this expression is $C_{G\gamma_5}(t=0)=0$ and is in accordance with the observation of item (1).

\begin{figure}[t]
    \includegraphics[width=0.9\linewidth]{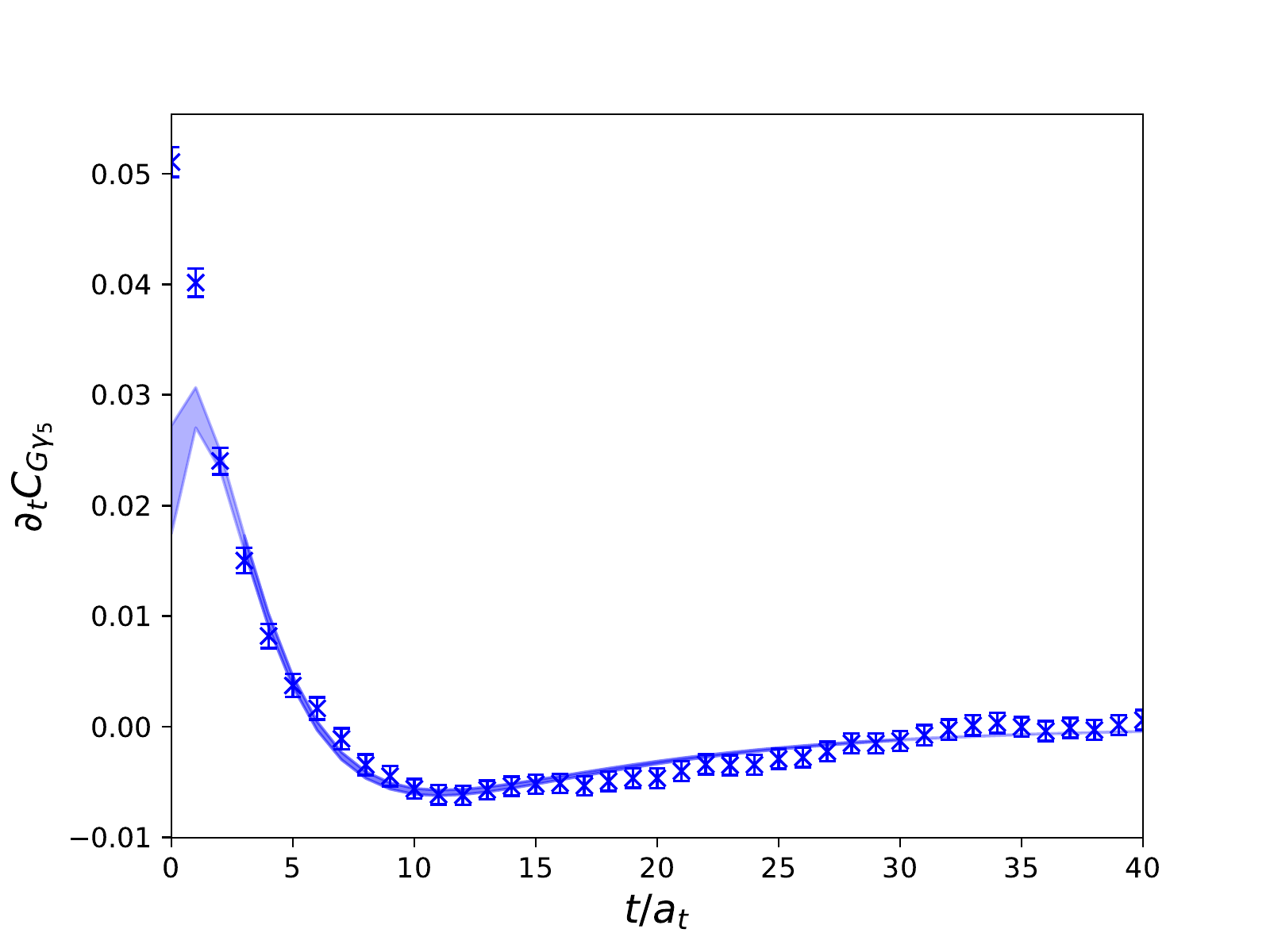}
    \caption{\label{fig:corr-eta-G}The temporal derivative of $C_{G\gamma_5}$ defined by Eq.~(\ref{eq:derivative}). The data points are lattice results and the shaded band illustrates the fitting results using Eq.~(\ref{eq:gc_general}). The signal to noise ratio of $\partial_tC_{G\gamma_5}$ is much better than that of $C_{G\gamma_5}$ in Fig.~\ref{fig:corr-eta-G-origin}.}
\end{figure}

In order to understand the almost constant error of $C_{G\gamma_5}(t)$, we consider its variance~\cite{Endres:2011mm}
\begin{equation}
    \delta^2 C_{ G\gamma_5}(t)\equiv \langle \mathcal{O}_G^2(t)\mathcal{O}_{\gamma_5}^2 (0) \rangle -C_{ G\gamma_5}^2 (t).
\end{equation}
The first term on the right-hand side can be viewed as a correlation function of the operator $\mathcal{O}^2_{\gamma_5}$ and $\mathcal{O}_G^2$, both of which have the vacuum quantum number $0^{++}$ (in the continuum limit) and are expected to have nonzero vacuum expectation values $\langle \mathcal{O}^2_{\gamma_5}\rangle\ne 0$ and $\langle \mathcal{O}_G^2\rangle\ne0$. Thus we have
\begin{equation}\label{eq:cg5}
    \delta^2 C_{G\gamma_5 }(t)= \langle \overline{\mathcal{O}_G^2}(t)\overline{\mathcal{O}_{\gamma_5}^2}(0) \rangle-C_{G\gamma_5}^2 (t)+\langle \mathcal{O}_G^2\rangle\langle\mathcal{O}^2_{\gamma_5}\rangle
\end{equation}
where $\overline {\mathcal{O}_i^2}(t)\equiv \mathcal{O}_i^2(t)-\langle \mathcal{O}^2_i\rangle$. The almost constant error of $C_{G\gamma_5}(t)$ implies that the constant term $\langle\mathcal{O}^2_{\gamma_5}\rangle \langle \mathcal{O}_G^2\rangle$ is large and dominate the variance. This is consistent with the argument in Ref.~\cite{Endres:2011mm} that the variance of a correlation function is dominated by the possible lowest state, which corresponds to the vacuum state with $E_{\mathrm{vac}}=0$ in our case. This motivates us to consider the temporal derivative of $C_{G\gamma_5}(t)$, namely,
\begin{equation}\label{eq:derivative}
    \partial_t C_{G\gamma_5}(t)=\frac{1}{2a_t}\left(C_{G\gamma_5 }(t+a_t)-C_{G\gamma_5}(t-a_t)\right),
\end{equation}
such that the constant term in $C_{G\gamma_5}(t)$ and its constant variance can be canceled. This is surely the case. We plot $\partial_t C_{G\gamma_5}(t)$ in Fig.~\ref{fig:corr-eta-G}, where one can see that $\partial_t C_{G\gamma_5}(t)$ goes to zero when $t$ is large and  its relative error is much smaller.

\begin{figure}[t]
    \includegraphics[width=0.9\linewidth]{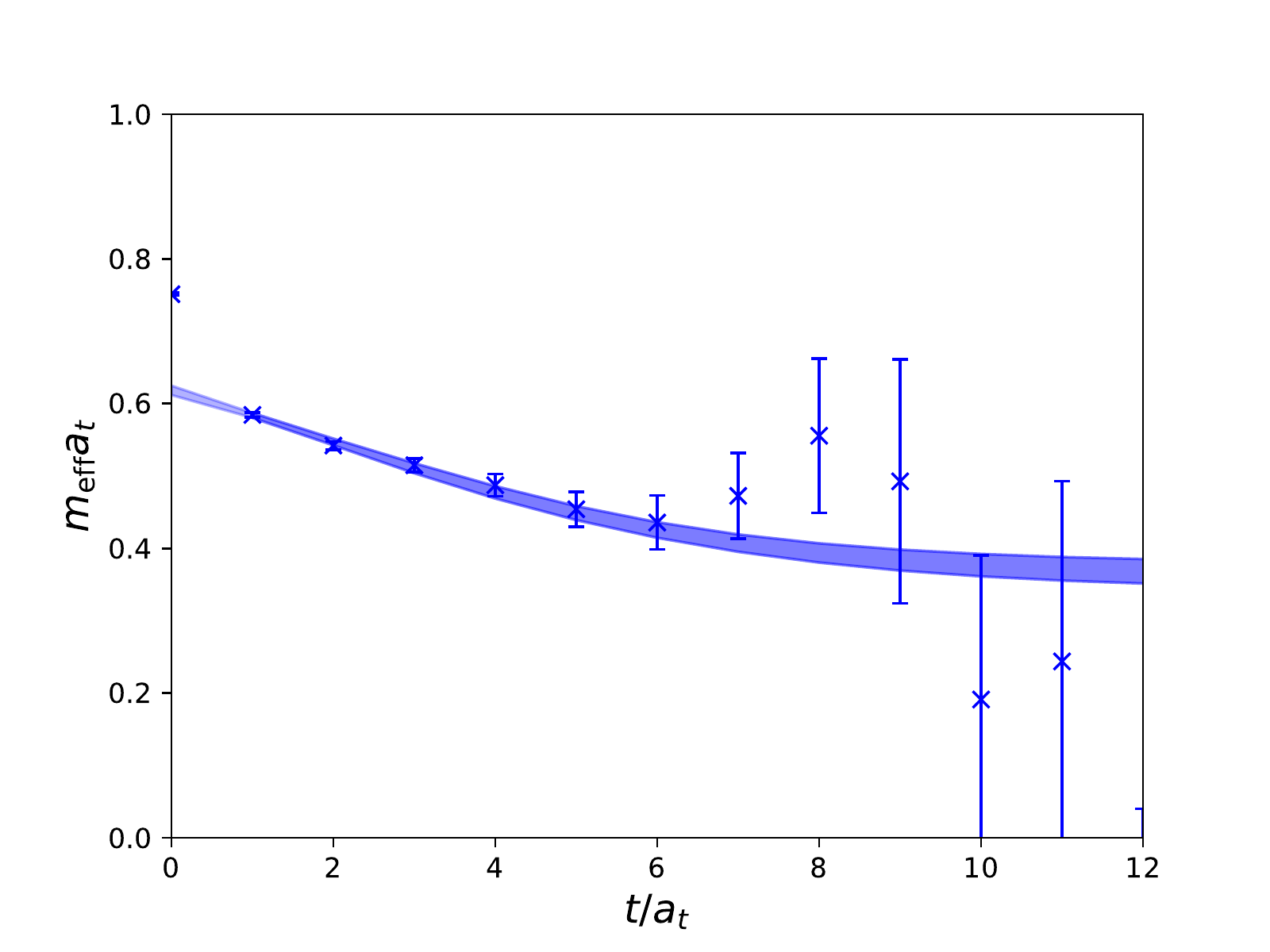}
    \caption{\label{fig:geffm} The effective mass of the pseudoscalar glueball. The shaded band illustrates the fitting results using Eq.~(\ref{eq:gg-gamma5}) with two mass terms.}
\end{figure}

In order for the mixing energies $x_1$ and $y_1$ to be extracted by using Eq.~(\ref{eq:gc_general}), one has to know the parameters $m_1$, $m_3$, $m_4$, $m_{\eta}$, $m_G$, $m_{G^*}$, $Z_G$, $Z_{G^{*}}$ and $Z_{\gamma_5,1}$, which, based on the assumptions of Eq.~(\ref{eq:create}), are encoded in the correlation functions $C_{\gamma_5\gamma_5}(t)$ and $C_{GG}(t)$ as
\begin{eqnarray}\label{eq:gg-gamma5}
    C_{GG}(t) &=& \sum\limits_i Z_{G_i}\left(e^{-m_{G_i}t}+e^{-m_{G_i}(T-t)}\right)\nonumber\\
    C_{\gamma_5\gamma_5}(t) &=& \sum\limits_i Z_{\gamma_5,i}\left(e^{-m_{\eta_i}t}+e^{-m_{\eta_i}(T-t)}\right)\nonumber\\
    &\approx& Z_{\gamma_5,1}\left(e^{-m_{\eta}t}+e^{-m_{\eta}(T-t)}\right)
\end{eqnarray}
where we take $i=1,2$ for $C_{GG}(t)$ and therefore $G_{1,2}$ refer to $G$ and $G^*$. It should be noted that the second state $|G^*\rangle$ should be considered even though we have built the optimal operator $\mathcal{O}_G$ based on a large operator set (seen in Appendix), it turns out that there is still a substantial contribution of higher states in $C_{GG}$. To manifest this, we plot the effective mass function $m_\mathrm{eff}(t)$ of $C_{GG}(t)$ in Fig.~\ref{fig:geffm} by the definition in Eq.~(\ref{eq:geffm}), where one can see that the ground state glueball $G$ has not saturated $C_{GG}(t)$ before the signals are undermined by errors. With this observation, we add the second term relevant to $G^*$ to the first equation in Eq.~(\ref{eq:gg-gamma5}) but do not consider its physical meaning.

%%%%%%%%%%%%%%%%%%%%%%%%%%%%%%%%%%%%%%%%%%%%%%%%%%%%%%%%%%%%%%%%%%%%%%%%%%
%%%%%%%%%%%%%%%%%%%%%%%%%%%%%%%%%%%%%%%%%%%%%%%%%%%%%%%%%%%%%%%%%%%%%%%%%%

\begin{table*}[t]
    \caption{Ground state mass and mixing angle fitted from operators with $\Gamma=\gamma_5$ and $\Gamma=\gamma_4\gamma_5$ on the ensemble. Values of $m_\eta$ are the same in both cases because the value is derived from a combined fit to $C_{\gamma_5\gamma_5}$ and $C_{(\gamma_4\gamma_5)(\gamma_4\gamma_5)}$. $\chi^2$ are obtained from the fitting results of $C_{G\Gamma}$.}
    \label{tab:fit}
    \begin{ruledtabular}
        \begin{tabular}{lcccc|cccc}
            $\Gamma$           & $[t_l,t_h]_{\Gamma}$ & $[t_l,t_h]_{GG}$ & $[t_l,t_h]_{G\Gamma}$ & $\chi^2/\mathrm{dof}$ & $m_{\eta}a_t$ & $m_{G}a_t$ & $|x_1|a_t$ & $|\theta|$       \\\hline
            $\gamma_5$         & [9, 30]              & [1, 14]          & [3, 30]               & $0.96$                & $0.10358(84)$    & $0.3607(75)$   & $0.0155(22)$    & $3.46(46)^\circ$ \\
            $\gamma_4\gamma_5$ & [5, 30]              & [1, 14]          & [0, 30]               & $0.15$                & $0.10358(84)$    & $0.3607(75)$   & $0.0112(55)$     & $2.5(1.2)^\circ$ \\
        \end{tabular}
    \end{ruledtabular}
\end{table*}

\begin{figure}[ht]
    \includegraphics[width=0.9\linewidth]{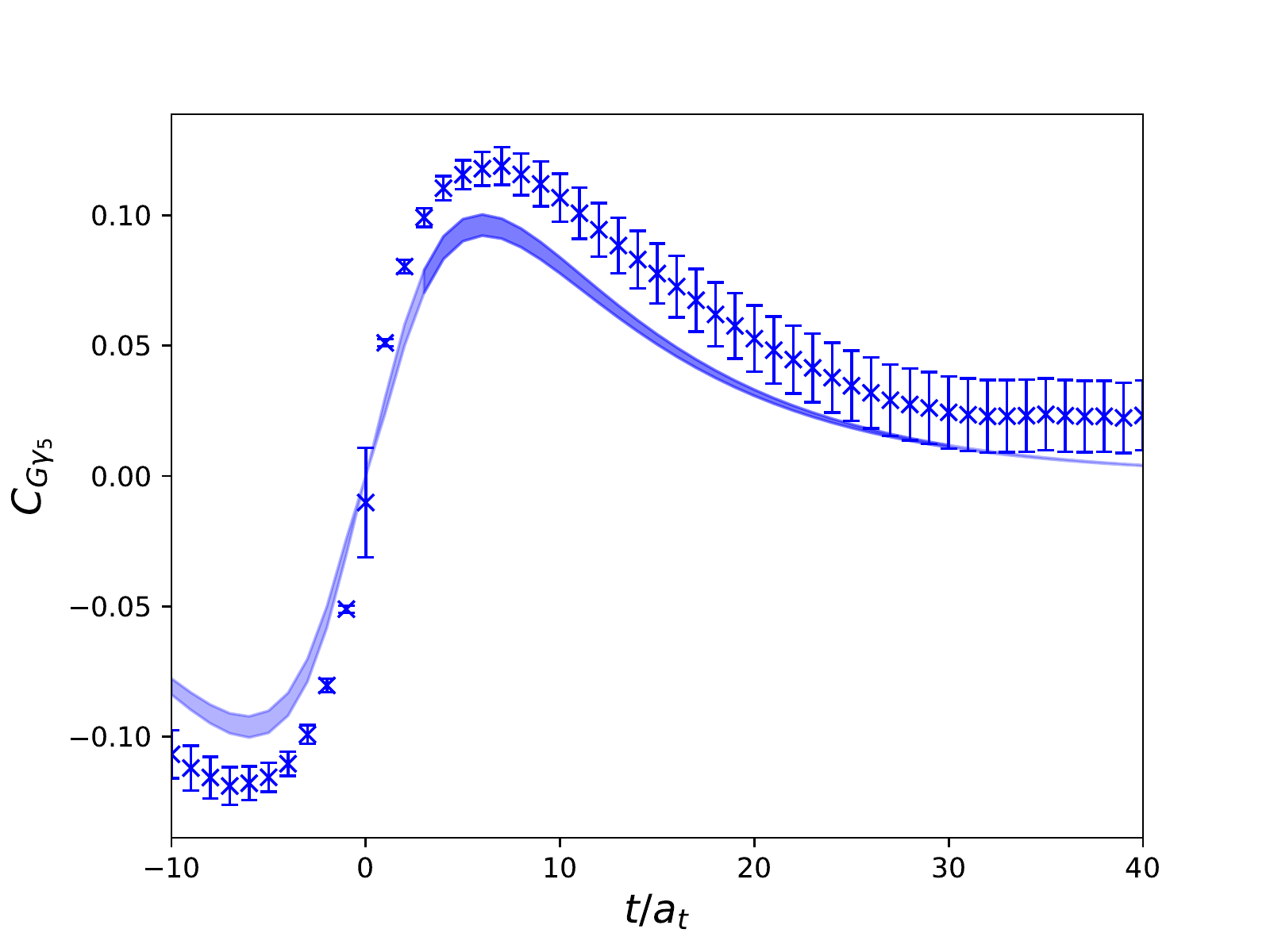}
    \caption{\label{fig:corr-eta-G-fold}$C_{G\gamma_5}$, and used $\mathcal{T}$-odd property to fold the data for every gauge configuration. The blue band shows the fitted results. The band doesn't perfectly match the data point because we just fitted temporal derivative $\partial_t C_{G\gamma_5}$ instead of correlation function $C_{G\gamma_5}$, so we dropped some constant in the fitting result. The $\mathcal{T}$-odd property makes the constant negative in $-t$ direction. The x-axis is shifted by 10 to show the near-zero behavior.}
\end{figure}

Since the mixing effect is expected to be small (our final results confirm this), we take the assumptions $m_1\approx m_\eta$, $m_3\approx m_G$ and $m_4\approx m_{G^*}$. In the data analysis procedure, we first rearrange the $N_\mathrm{conf}$ measurements into $139$ bins with each bin including $50$ measurements, and then we perform the one-eliminating jackknife analysis on these data bins. For each time of jackknife resampling, we first extract $m_{\eta}$, $m_G$, $m_{G^*}$, $Z_G$, $Z_{G^{*}}$, $Z_{\gamma_4\gamma_5,1}$ and $Z_{\gamma_5,1}$ from $C_{GG}(t)$, $C_{\gamma_5\gamma_5}(t)$ and $C_{(\gamma_4\gamma_5)(\gamma_4\gamma_5)}(t)$ in the fixed time windows $[t_l,t_h]_{GG}=[1,14]$, $[t_l,t_h]_{\gamma_5}=[9,30]$ and $[t_l,t_h]_{\gamma_4\gamma_5}=[5,30]$, as shown in Table~\ref{tab:fit}. Then we feed these parameters to $\partial_t C_{G\gamma_5}(t)$ to determine the parameters $x_1$ and $y_1$. The final results of $m_\eta$, $m_G$ and $|x_1|$ with jackknife errors are obtained to be
\begin{eqnarray}
    m_\eta a_t&=&0.10358(84),\nonumber\\
    m_{G} a_t&=&0.3607(75),\nonumber\\
    |x_1| a_t&=&0.0155(22).
\end{eqnarray}
Note that the definitions in Eqs.~(\ref{eq:create}), (\ref{eq:gcouple}), (\ref{eq:etacouple}) are up to a plus or minus sign, we can only determine the absolute value $|x_1|$ of $x_1$. The parameters of the fitting procedure and fitting results are collected in Table~\ref{tab:fit}. The goodness of the fit of Eq.~(\ref{eq:gc_general}) to $\partial_t C_{G\gamma_5}$ using the function is reflected by the $\chi^2/\mathrm{dof}=0.96$ in the fitting window $[t_l,t_h]_{G\gamma_5}=[3,30]$, and is also illustrated in Fig.~\ref{fig:corr-eta-G} by the shaded curve. In mean time, making use of the $\mathcal{T}$-odd property of $C_{G\gamma_5}(t)$, we average the $t<\frac{T}{2}$ part and $t>\frac{T}{2}$ part and find the errors can be reduced drastically around $t=0$ except for $C_{G\gamma_5}(t=0)$, as shown in Fig.~\ref{fig:corr-eta-G-fold}, where the function of Eq.~(\ref{eq:gc_general}) with fitted parameters is also plotted with shaded curves. It is seen that the function describes the $t$-dependence of $C_{G\gamma_5}(t)$ very well up to an unknown constant term with opposite signs for $t<\frac{T}{2}$ and $t>\frac{T}{2}$.

Since $|x_1|$ is much smaller than $m_G-m_\eta$, to the lowest order of the perturbation theory, we can estimate the mixing angle $\theta$ of $\eta$ and the ground state glueball $G$ as
\begin{equation}
    |\theta|\approx \sin |\theta|\approx \frac{|x_1|}{m_G-m_\eta}=3.46(46)^\circ.
\end{equation}

\subsection{The $\Gamma=\gamma_4\gamma_5$ case}
As a cross-check, we also carry a similar calculation by using the $\Gamma=\gamma_4\gamma_5$ for the interpolation field operator of $\eta$. The corresponding correlation functions $C_{(\gamma_4\gamma_5)(\gamma_4\gamma_5)}(t)$ and $C_{G(\gamma_4\gamma_5)}(t)$ are calculated using Eq.~(\ref{eq:corrs}). In contrast to the case of $\Gamma=\gamma_5$, the correlation function $C_{G(\gamma_4\gamma_5)}(t)$ does not go to zero when $t\to 0$. This is similar to the study of mixing of the pseudoscalar charmonium and the pseudoscalar glueball~\cite{Zhang:2021xvl}, which can be explained following the same logic that the QCD $\mathrm{U_A(1)}$ anomaly may play an important role here. Obviously, the operator $\mathcal{O}_{\gamma_4\gamma_5}$ has the same operator structure as the temporal component of the isoscalar axial vector current
$j^5_\mu(x)=\frac{1}{\sqrt{2}}\left[\bar{u}(x)\gamma_\mu\gamma_5 u(x)+\bar{d}(x)\gamma_\mu\gamma_5 d(x)\right]$, which satisfies the anomalous axial vector relation
\begin{equation}\label{eq:u1a}
    \partial_\mu j^5_\mu(x) = 2m_q j^5(x) + \sqrt{2} q(x),
\end{equation}
where $j^5(x)=\frac{1}{\sqrt{2}}\left[\bar{u}(x)\gamma_5 u(x)+\bar{d}(x)\gamma_5 d(x)\right]$ is the pseudoscalar density and $q(x)=\frac{g^2}{32\pi^2} \epsilon^{\alpha\beta\rho\sigma} G_{\alpha\beta}^a G^a_{\rho\sigma}$ is the anomalous term stemming from the $\mathrm{U_A(1)}$ anomaly with $g$ being the strong coupling constant and $G_{\alpha\beta}^a$ being the strength of color fields. Since $j^5(x)$ also has the same structure as $\mathcal{O}_{\gamma_5}$, based on the assumption in Eq.~(\ref{eq:create}) we expect
\begin{equation}
    m_{G_i}^2 f_{G_i} =\langle 0|\partial_\mu j^5_\mu(0)|G_i\rangle \approx \sqrt{2} \langle 0|q(0)|G_i\rangle
\end{equation}
where $f_{G_i}$ is the {\it decay constant} of the pseudoscalar glueball $G_i$. Accordingly we have
\begin{equation}
    \langle 0|j_4^5(0)|G_i(\vec{p}=0)\rangle \approx \frac{\sqrt{2}}{m_{G_i}} \langle 0|q(0)|G_i(\vec{p}=0)\rangle.
\end{equation}
%%%%%%%%%%%%%%%%%%%%%%%%%%%%%%%%%%%%%%%%%%%%%%%%%%%%%%%%%%%%%%%%%%%%%%%%%%%
\begin{figure}[t]
    \includegraphics[width=0.9\linewidth]{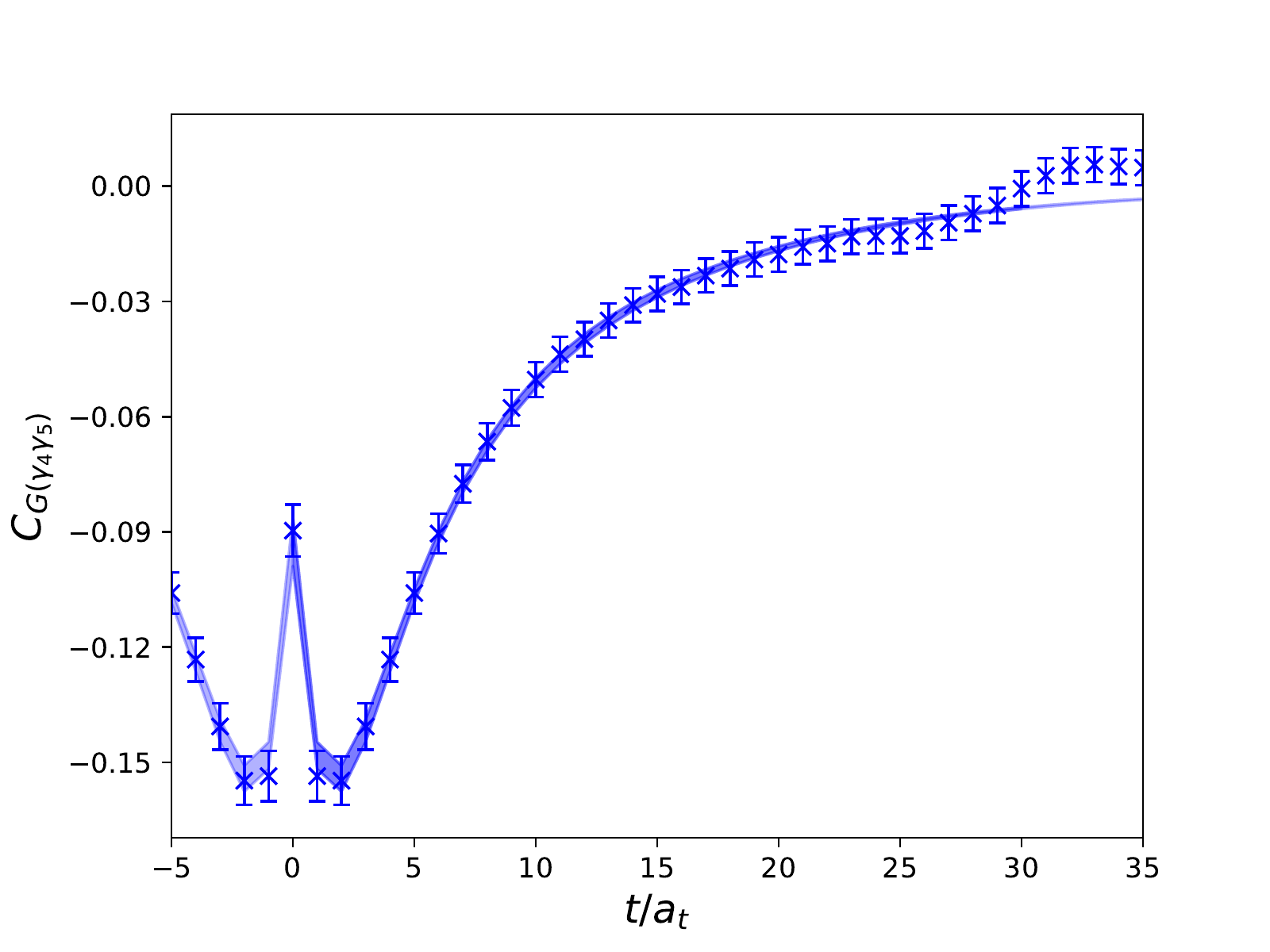}
    \caption{\label{fig:corr-eta-G-45} The correlation function of $\eta-G$ with $\Gamma=\gamma_4\gamma_5$, called $C_{G(\gamma_4\gamma_5)}$. The shaded band illustrates the fitting results using Eq.~(\ref{eq:cg45}). The fit window starts from $t=0$ in order to match the near-zero behavior of the correlator. The x-axis is shifted by 5 to show the near-zero behavior.}
\end{figure}
%%%%%%%%%%%%%%%%%%%%%%%%%%%%%%%%%%%%%%%%%%%%%%%%%%%%%%%%%%%%%%%%%%%%%%%%%%%
Previous lattice studies show that pseudoscalar states can be accessed by the operator $q(x)$~\cite{Chen:2005mg,Chowdhury:2014mra}, thus the nonzero matrix element $\langle 0|q(0)|G_i\rangle$ implies the coupling $\langle 0| \mathcal{O}_{\gamma_4\gamma_5}|G_i\rangle \neq 0$. If we insist the relation $\mathcal{O}_G^\dagger|0\rangle=\sum\limits_{i\neq 0}\sqrt{Z_{G_i}}|G_i\rangle$ still holds, then the correlation function $C_{G(\gamma_4\gamma_5)}(t)$ can be parametrized as
\begin{eqnarray}\label{eq:cg45}
    C_{G(\gamma_4\gamma_5)}(t)&\approx& A e^{-m_3 t}\nonumber\\
    &+&\sqrt{Z_G Z_{\gamma_4\gamma_5,1}}\frac{x_1 }{m_\eta-m_G} \left(e^{-m_1 t}-e^{-m_3 t}\right)\nonumber\\
    &+&\sqrt{Z_{G^*} Z_{\gamma_4\gamma_5,1}}\frac{y_1 }{m_{\eta}-m_{G^*}} \left(e^{-m_1 t}-e^{-m_4 t}\right)\nonumber\\
    &+& (t\to (T-t)~~ \mathrm{terms}),
\end{eqnarray}
which is similar to Eq.~(\ref{eq:gc_general}) apart from the first term due to nonzero coupling $\langle 0|q(0)|G_i\rangle$ (here we ignore the contribution of excited $\eta$ states). Alongside the correlation function
\begin{equation}\label{eq:gamma45-corr}
    C_{(\gamma_4\gamma_5)(\gamma_4\gamma_5)}(t)=\sum\limits_i Z_{\gamma_4\gamma_5,i}\left(e^{-m_{\eta_i}t}+e^{-m_{\eta_i}(T-t)}\right)
\end{equation}
and the first equation of Eq.~(\ref{eq:gg-gamma5}), we carry out a similar jackknife data analysis to the case of $\Gamma=\gamma_5$ and get the mixing energy $|x_1|a_t=0.0112(55)$ and the corresponding mixing angle $\theta_1=2.5(1.2)^\circ$, which are consistent with the results of $\Gamma=\gamma_5$ case but have larger errors. The parameters of the fitting procedure and the fit results are also listed in Table~\ref{tab:fit} for comparison. In Fig.~\ref{fig:corr-eta-G-45}, the colored curves with error bands are plotted using the fitted parameters. It is seen that the function forms mentioned above also describe the data very well. The fitted results are shown in Table~\ref{tab:fit} and can be compared with the $\gamma_5$-case directly. On this ensemble, it is clear that the results of the two cases are compatible with each other within errors.
%%%%%%%%%%%%%%%%%%%%%%%%%%%%%%%%%%%%%%%%%%%%%%%%%%%%%%%%%%%%%%%%%%%%%%%%%%%%%%%%%%%%%%%%%%%%%
% \begin{figure}[t]
%     \includegraphics[width=0.9\linewidth]{C_etac_eta.pdf}
%     \caption{\label{fig:corr-etac-eta}Correlation functions of $\eta-\eta_c$, called $C_{\Gamma\Gamma_c}$. Different colors indicate different source $\Gamma_c$ and sink $\Gamma$ insertions.}
% \end{figure}
%%%%%%%%%%%%%%%%%%%%%%%%%%%%%%%%%%%%%%%%%%%%%%%%%%%%%%%%%%%%%%%%%%%%%%%%%%%%%%%%%%%%%%%%%%%%%

%%%%%%%%%%%%%%%%%%%%%%%%%%%%%%%%%%%%%%%%%%%%%%%%%%%%%%%%%%%%%%%%%%%%%%%%%%%%%%%%%%%%%%%%%%%%%
\begin{figure}[t]
    \includegraphics[width=0.9\linewidth]{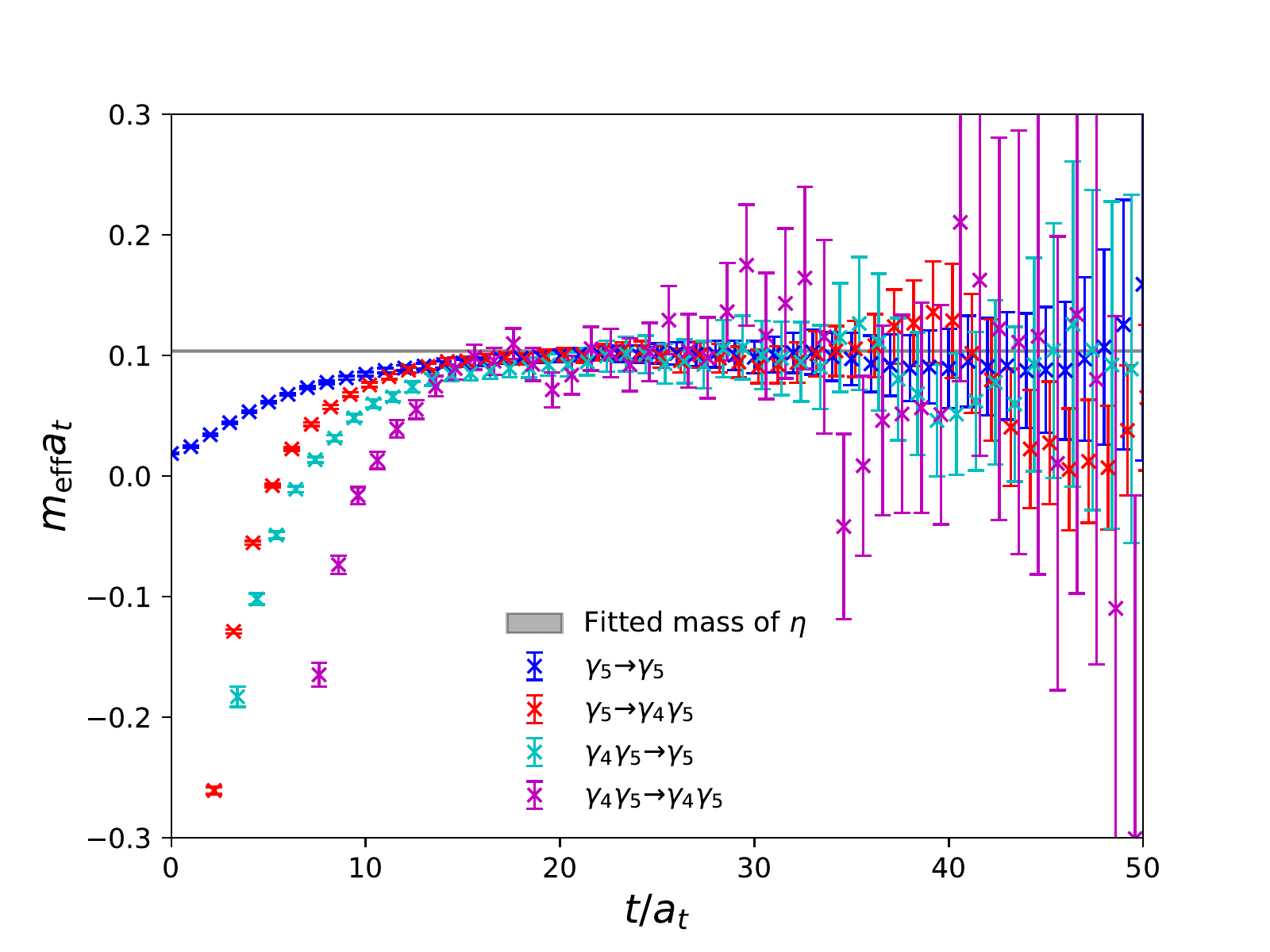}
    \caption{\label{fig:mass-etac-eta}Effective mass functions of $C_{\Gamma\Gamma_c}$ defined by Eq.~(\ref{eq:geffm}). Different colors indicate different source $\Gamma_c$ and sink $\Gamma$ insertions. The gray band shows the fitted result of $m_\eta a_t$ with error, and we can see all four functions have plateaus near $m_\eta a_t$ at large $t$.}
\end{figure}
%%%%%%%%%%%%%%%%%%%%%%%%%%%%%%%%%%%%%%%%%%%%%%%%%%%%%%%%%%%%%%%%%%%%%%%%%%%%%%%%%%%%%%%%%%%%%

\section{Discussion and Summary}\label{sec:summary}
We generate a large gauge ensemble with $N_f=2$ degenerated $u,d$ dynamical quarks on a $16^3\times 128$ anisotropic lattice with the anisotropy $\xi=a_s/a_t\approx 5.3$. Based on the experimental observation that the differences of mass squares of the heavy-light vector and pseudoscalar mesons are insensitive to quark masses, we propose a new scale setting scheme that the temporal lattice spacing $a_t$ can be estimated by this quantity. In practice, we use the value $m_V^2-m_{PS}^2= 0.568(8)~\mathrm{GeV}^2$, which is a least squares fitting of $n\bar{n}$, $s\bar{n}$, $s\bar{s}$, $c\bar{n}$ and $c\bar{s}$ mesons with $n$ referring to the light $u,d$ quarks. Our $u,d$ quark mass parameter corresponds to a pion mass $m_\pi\approx 350$ MeV. We calculate the perambulators of $u,d$ quarks and the valence charm quark on our ensemble using the distillation method where the quark annihilation diagrams can be calculated efficiently.

We calculate the mass of the isoscalar pseudoscalar meson $\eta$ to be $m_\eta=714(6)(16)$ MeV. The error in the latter bracket is introduced by the estimation of lattice spacing in Sec. \ref{sec:setup}. This mass value can be matched, through the Witten-Veneziano relation, to the $\mathrm{SU(3)}$ flavor singlet pseudoscalar meson mass $m_{\eta_1}\approx 981$ MeV, which is in good agreement with the $\eta'$ mass $m_{\eta'}=958$ MeV if the $\eta-\eta'$ mixing is considered.

The mixing of $\eta$ and the pseudoscalar glueball is investigated for the first time on the lattice. From the correlation function of the glueball operator and $\eta$ operator, the mixing energy and the mixing angle are determined to be $|x_1|=107(15)(2)$ MeV and $|\theta|=3.46(46)^\circ$ given the pseudoscalar glueball mass $m_G\approx 2.5$ GeV. This mixing angle is so tiny that the mixing effects can be neglected when the properties of $\eta$ (and also $\eta'$ in the physical $N_f=3$ case) are explored.

With perambulators of light and charm quarks at hand, we also checked correlation functions of $\eta$ and $\eta_c$, which are contributed only by annihilation diagrams. The $\Gamma$ insertion are chosen to be $\gamma_5$ and $\gamma_4\gamma_5$. The signal-to-noise ratios of these correlation functions are fairly good, and the effective masses of them defined by Eq.~(\ref{eq:geffm}) are presented in Fig~\ref{fig:mass-etac-eta}, where the horizontal gray line illustrates the $\eta$ mass in the lattice unit. Albeit the different behaviors at the small $t$ range, all the effective masses approach $m_\eta$ when $t$ is large. Because of the influence from excited states of $\eta$ and $\eta_c$, we cannot fit the curve by the simplification above, and optimized operators for these states are required to get better results. This kind of exploration can be performed in the future.

To summarize, we find that there is little mixing between the flavor singlet pseudoscalar ($\eta$ is our case) and the pseudoscalar glueball. The topology of the QCD vacuum plays a crucial role in the origin of $\eta$ mass due to the QCD $\mathrm{U_A(1)}$ anomaly. This is in accordance with the common theoretical considerations~\cite{Bass:2018xmz}.

\vspace{0.5cm}
\begin{acknowledgements}
    This work is supported by the National Key Research and Development Program of China (No. 2020YFA0406400), the Strategic Priority Research Program of Chinese Academy of Sciences (No. XDB34030302), and the National Natural Science Foundation of China (NNSFC) under Grants No. 11935017, No. 11775229, No. 12075253, No. 12070131001 (CRC 110 by DFG and NNSFC) and No. 12175063. The Chroma software system~\cite{Edwards:2004sx} and QUDA library~\cite{Clark:2009wm,Babich:2011np} are acknowledged. The computations were performed on the HPC clusters at Institute of High Energy Physics (Beijing) and China Spallation Neutron Source (Dongguan), and the CAS ORISE computing environment.
\end{acknowledgements}

\section*{APPENDIX}\label{sec:appendix}
\subsection*{A1. Glueball operator construction}
\begin{figure}[t]
    \centering
    \includegraphics[width=0.9\linewidth]{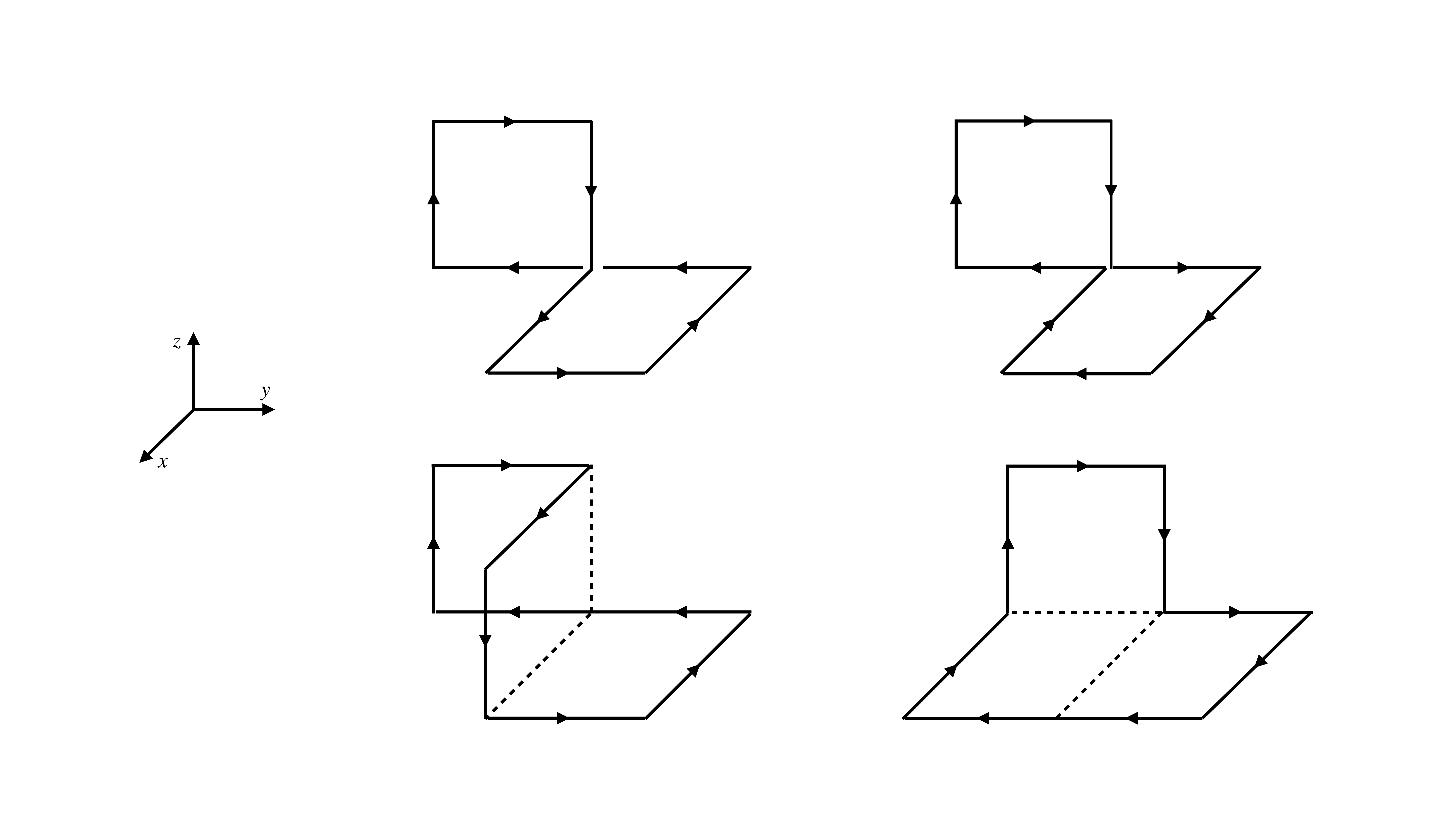}
    \caption{\label{fig:O_G}Wilson loop prototypes used to construct the pseudoscalar glueball operator~\cite{Morningstar:1999rf,Chen:2005mg}.}
\end{figure}
The interpolation field operators for the pseudoscalar glueball are built based on the four prototypes of Wilson loops shown in Fig.~\ref{fig:O_G}. The Wilson loops of each prototype are calculated using smeared gauge links. We adopt six different schemes to smear gauge links, which are different combinations of single link smearing and double link smearing~\cite{Morningstar:1999rf,Chen:2005mg}. The spatial symmetry group on the lattice is the octahedral group $O$ whose irreducible representation $A_1$ corresponds to the spin $J=0,4,\ldots$ in the continuum limit. It is expected the glueball of $J=4$ is higher than the $J=0$ state in mass, so we can build operators in the $A_1^{-+}$ representation to couple with the ground state pseudoscalar glueball. Let $W_\alpha(\mathbf{x},t)$ be one prototype of the Wilson loop under a specific smearing scheme, then the $A_1^{-+}$ operator in the rest frame of a glueball can be obtained by
\begin{equation}
    \phi_\alpha (t)=\sum\limits_{\mathbf{x}}\sum\limits_{R\in O} c_R^{A_1}\left[R\circ W_\alpha(\mathbf{x},t)-\mathcal{P}R\circ W_\alpha(\mathbf{x},t)\mathcal{P}^{-1}\right]
\end{equation}
where $R\circ W_\alpha$ refers to differently oriented Wilson loops after one of the 24 elements of $O$ ($R$) operated on $W_\alpha$, $\mathcal{P}$ is spatial reflection operation and $c_R^{A_1}$ are the combinational coefficients for the $A_1$ representation. Thus we obtain a $A_1^{-+}$ operator set $\{\phi_\alpha(t),\alpha=1,2,\ldots, 24\}$ based on the four prototypes and six smearing schemes. In order to get an optimized operator $O_G$ that couples most to the ground state glueball, we first calculate the correlation matrix of the operator set,
\begin{equation}
    C_{\alpha\beta}(t)=\frac{1}{T}\sum\limits_{\tau}\langle \phi_\alpha(t+\tau)\phi_\beta(\tau)\rangle,
\end{equation}
and then solve the generalized eigenvalue problem
\begin{equation}
    C_{\alpha\beta}(t_1)w_\beta=\lambda C_{\alpha\beta}(t_0)w_\beta
\end{equation}
to get the eigenvector $w_\alpha$ of the largest eigenvalue $\lambda$, which serves as the combinational coefficients of $O_G$, namely,
\begin{equation}
    O_G(t)=w_\alpha \phi_\alpha(t).
\end{equation}
In practice, we take $t_0=1$ and $t_1=2$ for the optimized operator coupling more to the ground state pseudoscalar glueball.

\bibliography{ref}

\end{document}